\documentclass[fleqn,usenatbib]{mnras}

\usepackage{newtxtext,newtxmath}

\usepackage[T1]{fontenc}

\DeclareRobustCommand{\VAN}[3]{#2}
\let\VANthebibliography\thebibliography
\def\thebibliography{\DeclareRobustCommand{\VAN}[3]{##3}\VANthebibliography}

\usepackage{graphicx}	
\usepackage{amsmath}	
\usepackage{subcaption}
\usepackage{orcidlink}
\usepackage{gensymb}
\usepackage{multicol}
\usepackage{booktabs,caption}
\usepackage[flushleft]{threeparttable}

\newcommand{\TESS}{{\it TESS}}
\newcommand{\tess}{{\it TESS}}
\newcommand{\PLATO}{{\it PLATO}}
\newcommand{\plato}{{\it PLATO}}
\newcommand{\gaia}{{\it Gaia}}

\newcommand{\NGTS}{{\it NGTS}}
\newcommand{\ngts}{{\it NGTS}}

\newcommand{\harps}{{\it HARPS}}

\newcommand{\feros}{{\it FEROS}}

\newcommand{\coralie}{{\it CORALIE}}

\newcommand{\cheops}{{\it CHEOPS}}
\newcommand{\CHEOPS}{{\it CHEOPS}}

\newcommand{\trappist}{{\it TRAPPIST}}

\newcommand{\speculoos}{{\it SPECULOOS}}

\newcommand{\kms}{km\,s$^{-1}$}

\newcommand{\mstar}{\mbox{M$_{\star}$}}
\newcommand{\rstar}{\mbox{R$_{\star}$}}

\newcommand{\mjup}{\mbox{M\textsubscript{J}}}
\newcommand{\rjup}{\mbox{R\textsubscript{J}}}
\newcommand{\msun}{\mbox{M$_{\odot}$}}
\newcommand{\rsun}{\mbox{R$_{\odot}$}}

\newcommand{\rearth}{R$_{\oplus}$}

\newcommand{\teff}{T\textsubscript{eff}}

\newcommand{\logg}{$\log g$}

\newcommand{\mone}{\mbox{M$_\text{A}$}}
\newcommand{\rone}{\mbox{R$_\text{A}$}}
\newcommand{\mtwo}{\mbox{M$_\text{B}$}}
\newcommand{\rtwo}{\mbox{R$_\text{B}$}}

\newcommand{\vsini}{$v \sin i_\star$}   
 
\newcommand{\vmic}{$v_{\rm mic}$}
\newcommand{\vmac}{$v_{\rm mac}$}

\newcommand{\met}{[M/H]}

\newcommand{\logrhk}{log\,R$^\prime_\mathrm{HK}$}

\newcommand{\TICID}{TIC\,238060327}
\newcommand{\OBJID}{\ngts-EB-7}

\newcommand{\logrhkval}{$-5.12\pm0.17$}
\newcommand{\secondarydurtot}{$12.7\pm0.2$}
\newcommand{\secondarydurfull}{$10.6\pm0.2$}
\newcommand{\Tnoughtocc}{$2459503.5\pm0.1$}
\newcommand{\phinoughtocc}{$0.0900\pm0.0006$}

\newcommand{\MBMsun}{$0.096^{+0.003}_{-0.004}$}
\newcommand{\MBMjup}{$102^{+4}_{-5}$}

\newcommand{\apodistance}{$1.13\pm0.06$}
\newcommand{\peridistance}{$0.189\pm0.009$}

\newcommand{\hostteff}{$5770\pm110$}
\newcommand{\hostlogg}{$4.17\pm0.15$}
\newcommand{\hostmet}{$0.26\pm0.12$}
\newcommand{\hostvsini}{$2.7\pm0.8$}
\newcommand{\hostrad}{$1.45^{+0.07}_{-0.06}$}
\newcommand{\isohostmass}{$1.13^{+0.06}_{-0.07}$}
\newcommand{\hostage}{$10\pm1$}

\newcommand{\alles}{\texttt{allesfitter}}
\newcommand{\paws}{\texttt{PAWS}}
\newcommand{\ariadne}{\texttt{astroARIADNE}}
\newcommand{\specmatch}{\texttt{specmatch-emp}}
\newcommand{\isochrones}{\texttt{isochrones}}

\newcommand{\Brr}{$0.0869_{-0.0011}^{+0.0012}$} 
\newcommand{\Brsuma}{$0.01098_{-0.00016}^{+0.00026}$} 
\newcommand{\Bcosi}{$0.0032_{-0.0021}^{+0.0028}$} 
\newcommand{\Bepoch}{$2459486.0746_{-0.0015}^{+0.0016}$} 
\newcommand{\Bperiod}{$193.35875\pm0.00034$} 
\newcommand{\Bfc}{$-0.83371\pm0.00076$} 
\newcommand{\Bfs}{$0.1389\pm0.0028$} 
\newcommand{\BK}{$4.6150\pm0.0096$} 
\newcommand{\hostldcqoneNGTS}{$0.375\pm0.010$} 
\newcommand{\hostldcqtwoNGTS}{$0.401\pm0.078$} 
\newcommand{\hostldcqoneTESSeight}{$0.3149\pm0.0083$} 
\newcommand{\hostldcqtwoTESSeight}{$0.379\pm0.069$} 
\newcommand{\hostldcqoneTESStwentynine}{$0.3152\pm0.0083$} 
\newcommand{\hostldcqtwoTESStwentynine}{$0.396\pm0.073$} 
\newcommand{\lnerrfluxTESSeight}{$-7.310\pm0.021$} 
\newcommand{\lnerrfluxTESStwentynine}{$-6.527\pm0.011$} 
\newcommand{\lnerrfluxNGTS}{$-7.130_{-0.085}^{+0.090}$} 
\newcommand{\lnjitterrvCORALIE}{$-3.99_{-0.69}^{+0.42}$} 
\newcommand{\lnjitterrvFEROS}{$-3.48\pm0.29$} 
\newcommand{\lnjitterrvHARPS}{$-4.27_{-0.25}^{+0.27}$} 
\newcommand{\baselinegpmaternthreetwolnsigmafluxTESSeight}{$-7.50_{-0.18}^{+0.20}$} 
\newcommand{\baselinegpmaternthreetwolnrhofluxTESSeight}{$0.07\pm0.27$} 
\newcommand{\baselinegpmaternthreetwolnsigmafluxTESStwentynine}{$-7.009\pm0.052$} 
\newcommand{\baselinegpmaternthreetwolnrhofluxTESStwentynine}{$-1.714\pm0.099$} 
\newcommand{\baselineoffsetrvCORALIE}{$78.4273\pm0.0089$} 
\newcommand{\baselineoffsetrvHARPS}{$78.3140\pm0.0064$} 
\newcommand{\baselineoffsetrvFEROS}{$78.418_{-0.011}^{+0.010}$} 
\newcommand{\BRstarovera}{$0.01011_{-0.00014}^{+0.00023}$} 
\newcommand{\BaoverRstar}{$98.9_{-2.2}^{+1.4}$} 
\newcommand{\BRcompanionovera}{$0.000878_{-0.000018}^{+0.000026}$} 
\newcommand{\BRcompanionRearth}{$13.66\pm0.65$} 
\newcommand{\BRcompanionRjup}{$1.219\pm0.058$} 
\newcommand{\BRcompanionRsun}{$0.125\pm0.006$}
\newcommand{\BaRsun}{$142.0\pm7.1$} 
\newcommand{\BaAU}{$0.661\pm0.033$} 
\newcommand{\Bi}{$89.82_{-0.16}^{+0.12}$} 
\newcommand{\Be}{$0.71436\pm0.00085$} 
\newcommand{\Bw}{$170.54_{-0.19}^{+0.20}$} 
\newcommand{\Bq}{$0.0952_{-0.0049}^{+0.0055}$} 
\newcommand{\BMcompanionMearth}{$35700_{-2700}^{+2900}$} 
\newcommand{\BMcompanionMjup}{$112.3_{-8.5}^{+9.2}$} 
\newcommand{\BMcompanionMsun}{$0.1072_{-0.0082}^{+0.0088}$} 
\newcommand{\Bbtra}{$0.138_{-0.090}^{+0.12}$} 
\newcommand{\BTtratot}{$10.06_{-0.10}^{+0.11}$} 
\newcommand{\BTtrafull}{$8.406_{-0.099}^{+0.092}$} 
\newcommand{\Bhostdensity}{$0.490_{-0.032}^{+0.022}$} 
\newcommand{\Bdensity}{$77_{-14}^{+17}$} 
\newcommand{\Bsurfacegravity}{$157400_{-9000}^{+6600}$} 
\newcommand{\BTeq}{$375.7\pm8.6$} 
\newcommand{\BdepthtrundilTESSeight}{$8.97_{-0.22}^{+0.25}$} 
\newcommand{\BdepthtrdilTESSeight}{$8.71_{-0.22}^{+0.24}$} 
\newcommand{\BdepthtrundilTESStwentynine}{$9.01\pm0.25$} 
\newcommand{\BdepthtrdilTESStwentynine}{$8.89\pm0.25$} 
\newcommand{\BdepthtrundilNGTS}{$9.17_{-0.29}^{+0.33}$} 
\newcommand{\BdepthtrdilNGTS}{$9.17_{-0.29}^{+0.33}$} 
\newcommand{\hostldcuoneTESSeight}{$0.424\pm0.078$} 
\newcommand{\hostldcutwoTESSeight}{$0.136\pm0.078$} 
\newcommand{\hostldcuoneTESStwentynine}{$0.444\pm0.083$} 
\newcommand{\hostldcutwoTESStwentynine}{$0.117\pm0.083$} 
\newcommand{\hostldcuoneNGTS}{$0.491\pm0.096$} 
\newcommand{\hostldcutwoNGTS}{$0.121\pm0.096$} 

\title[\OBJID]{\OBJID, an eccentric, long-period, low-mass eclipsing binary}

\author[T. Rodel et al.]{\parbox{\textwidth}{\Large
Toby Rodel$^{1}$\thanks{Email: trodel01@qub.ac.uk}\orcidlink{0009-0009-2175-72841},
Christopher. A. Watson$^{1}$\orcidlink{0000-0002-9718-3266},
Sol\`ene Ulmer-Moll$^{2,3}$\orcidlink{0000-0003-2417-7006},
Samuel Gill$^{4,5}$\orcidlink{0000-0002-4259-0155},
Pierre F. L. Maxted$^{6}$\orcidlink{0000-0003-3794-1317},
Sarah L. Casewell$^{7}$\orcidlink{0000-0003-2478-0120},
Rafael Brahm$^{8,9,10}$\orcidlink{0000-0002-9158-7315},
Thomas G Wilson$^{4,5}$\orcidlink{0000-0001-8749-1962},
Jean C. Costes$^{1}$,
Yoshi Nike Emilia Eschen$^{4,5}$\orcidlink{0009-0006-6397-2503},
Lauren Doyle$^{4,5}$\orcidlink{0000-0002-9365-2555},
Alix V. Freckelton$^{11}$\orcidlink{0009-0007-1053-0004},
Douglas R. Alves$^{12,13}$\orcidlink{0000-0002-5619-2502},
Ioannis Apergis$^{4,5}$\orcidlink{0009-0004-7473-4573},
Daniel Bayliss$^{4,5}$\orcidlink{0000-0001-6023-1335},
Francois Bouchy$^{2}$\orcidlink{0000-0002-7613-393},
Matthew R. Burleigh$^{7}$\orcidlink{0000-0003-0684-7803},
Xavier Dumusque$^{2}$\orcidlink{0000-0002-9332-2011},
Jan Eberhardt$^{14}$\orcidlink{0000-0003-3130-2768},
Jorge Fern\'andez Fern\'andez$^{4,5}$\orcidlink{0000-0002-1416-2188},
Edward Gillen$^{15}$\orcidlink{0000-0003-2851-3070},
Michael R. Goad$^{7}$\orcidlink{0000-0002-2908-7360},
Faith Hawthorn$^{4,5}$\orcidlink{0000-0002-8675-182X},
Ravit Helled$^{16}$\orcidlink{0000-0001-5555-2652},
Thomas Henning$^{14}$,
Katlyn L. Hobbs$^{1}$\orcidlink{0009-0004-4519-5080},
James S. Jenkins$^{17,13}$\orcidlink{0000-0003-2733-8725},
Andr\'es Jord\'an$^{9,8,10}$\orcidlink{0000-0002-5389-3944},
Alicia Kendall$^{7}$\orcidlink{0009-0006-0719-9229},
Monika Lendl$^{2}$\orcidlink{0000-0001-9699-1459},
James McCormac$^{4,5}$\orcidlink{0000-0003-1631-4170},
Ernst J.W. de Mooij$^{1}$\orcidlink{0000-0001-6391-9266},
Sean M. O'Brien$^{1}$\orcidlink{0000-0001-7367-1188},
Suman Saha$^{17,13}$\orcidlink{0000-0001-8018-0264},
Marcelo Tala Pinto$^{9,8}$\orcidlink{0009-0004-8891-4057},
Trifon Trifonov$^{18,19}$\orcidlink{0000-0002-0236-775X},
St\'{e}phane Udry$^{2}$\orcidlink{0000-0001-7576-6236},
Peter J. Wheatley$^{4,5}$\orcidlink{0000-0003-1452-2240}
}
\vspace{0.2cm}
\\
$^{1}$Astrophysics Research Centre, School of Mathematics and Physics, Queen’s University Belfast, Belfast, BT7 1NN, UK\\
$^{2}$Observatoire de Gen\`{e}ve, Universit\'{e} de Gen\`{e}ve, 51 Ch. des Maillettes, CH-1290 Sauverny, Switzerland\\
$^{3}$Space Research and Planetary Sciences, Physics Institute, University of Bern, Gesellschaftsstrasse 6, 3012 Bern, Switzerland\\
$^{4}$Department of Physics, University of Warwick, Gibbet Hill Road, Coventry CV4 7AL, UK\\
$^{5}$Centre for Exoplanets and Habitability, University of Warwick, Gibbet Hill Road, Coventry CV4 7AL, UK\\
$^{6}$Astrophysics Group, Keele University, Keele, Staffordshire ST5 5BG, UK\\
$^{7}$School of Physics and Astronomy, University of Leicester, Leicester LE1 7RH, UK\\
$^{8}$Millennium Institute of Astrophysics (MAS), Nuncio Monse\~{n}or S\' {o}tero Sanz 100, Providencia, Santiago, Chile\\
$^{9}$Facultad de Ingenier\'{ı}a y Ciencias, Universidad Adolfo Ib\'{a}\~{n}ez, Av. Diagonallas Torres, Pe\~{n}alol\'{e}n 2640, Santiago, Chile\\
$^{10}$Data Observatory Foundation, Eliodoro Y\'{a}\~{n}ez 2990, Providencia, Santiago,Chile\\
$^{11}$School of Physics \& Astronomy, University of Birmingham, Edgbaston, Birmingham B15 2TT, UK\\
$^{12}$Departamento de Astronom\'ia, Universidad de Chile, Casilla 36-D, Santiago, Chile\\
$^{13}$Centro de Excelencia en Astrofísica y Tecnologías Afines (CATA), Camino El Observatorio 1515, Las Condes, Santiago, Chile\\
$^{14}$Max-Planck-Institut f\"ur Astronomie, K\"onigstuhl 17, 69117 Heidelberg, Germany\\
$^{15}$Astronomy Unit, Queen Mary University of London, Mile End Road, London E1 4NS, UK\\
$^{16}$Department of Astrophysics, University of Zurich, Winterthurerstr. 190, 8057 Zurich, Switzerland\\
$^{17}$Instituto de Estudios Astrofísicos, Facultad de Ingeniería y Ciencias, Universidad Diego Portales, Av. Ejército Libertador 441, Santiago, Chile\\
$^{18}$Landessternwarte, Zentrum f\"ur Astronomie der Universt\"at Heidelberg, K\"onigstuhl 12, 69117 Heidelberg, Germany\\
$^{19}$Department of Astronomy, Sofia University St Kliment Ohridski, 5 James Bourchier Blvd, BG-1164 Sofia, Bulgaria\\
}

\date{Accepted XXX. Received YYY; in original form ZZZ}

\pubyear{2024}

\begin{document}
\label{firstpage}
\pagerange{\pageref{firstpage}--\pageref{lastpage}}
\maketitle

\begin{abstract}
Despite being the most common types of stars in the Galaxy, the physical properties of late M dwarfs are often poorly constrained. A trend of radius inflation compared to evolutionary models has been observed for earlier type M dwarfs in eclipsing binaries, possibly caused by magnetic activity. It is currently unclear whether this trend also extends to later type M dwarfs below the convective boundary. This makes the discovery of lower-mass, fully convective, M dwarfs in eclipsing binaries valuable for testing evolutionary models – especially in longer-period binaries where tidal interaction between the primary and secondary is negligible. With this context, we present the discovery of the \OBJID\,AB system, an eclipsing binary containing a late M dwarf secondary and an evolved G-type primary star. The secondary star has a radius of \BRcompanionRsun\,\rsun, a mass of \MBMsun\,\msun\ and follows a highly eccentric (e=\Be) orbit every \Bperiod\,days. This makes \OBJID\,AB the third longest-period eclipsing binary system with a secondary smaller than 200\,\mjup\ with the mass and radius constrained to better than 5\%. In addition, \OBJID\ is situated near the centre of the proposed LOPS2 southern field of the upcoming \plato\ mission, allowing for detection of the secondary eclipse and measurement of the companion's temperature. With its long-period and well-constrained physical properties - \OBJID\,B will make a valuable addition to the sample of M dwarfs in eclipsing binaries and help in determining accurate empirical mass/radius relations for later M dwarf stars.

\end{abstract}

\begin{keywords}
stars: binaries: eclipsing -- stars: fundamental properties -- stars: late type -- stars: low mass -- planets and satellites: detection -- planets and satellites: fundamental properties
\end{keywords}



\section{Introduction}
\label{section:intro}

In recent years late M dwarfs have become attractive targets for exoplanet surveys attempting to find temperate Earth-sized planets. The small radii and masses of M dwarfs means an Earth-sized planet produces deeper transits and induces a larger amplitude reflex motion on its host star relative to higher-mass stars, both of which allow for easier detection. Additionally, the cooler temperatures of late M dwarfs coupled to their small radii (and therefore lower luminosities) means that the `temperate zone' is much closer to the star than it would be around a solar-type star. Since closer-in planets are easier to detect with both transits and radial velocities, this makes detection of temperate (and thus potentially habitable) planets more likely around late M dwarfs than earlier type stars. These factors have led to large scale surveys such as MEarth \citep{Nutzman2009MEarth} and the Search for habitable Planets EClipsing ULtra-cOOl Stars \citep[\speculoos;][]{Sebastian2021SPECULOOS} being dedicated entirely to searching for planets around M dwarfs and other large surveys such as the Transiting Exoplanet Survey Satellite \citep[\tess;][]{ricker2015tess} prioritising nearby M dwarfs as targets. Such survey efforts have already led to exciting discoveries, including a compact system of 7 planets (up to 4 of which are potentially temperate) similar in size to Earth orbiting the M8V \citep{Costa2006} star \trappist-1 \citep{Gillon2017TRAPPIST1}.

In order to obtain accurate planetary parameters, the host star itself must first be accurately characterised. However, there has been a long-observed discrepancy between the theoretical masses and radii from evolutionary models and the measured masses and radii of M dwarfs in eclipsing binaries -- which was identified as early as \citet{Hoxie1970radiusinflation}. Studies such as \citet{Spada2013radiusinflation} and \citet{Parsons2018radiusinflation} have found M dwarf binary companions are often cooler and larger than predicted by models by a factor of 3-10\%. There is currently conflicting evidence for whether this `radius inflation problem' is confined to earlier type M dwarfs between 0.35\,\msun\ and 0.5\,\msun\ or whether it also extends to fully convective later type M dwarfs below 0.35\,\msun. \citet{Parsons2018radiusinflation} found that radius inflation was present either side of the convective boundary, while \citet{vonBoetticher2019EBLM5} instead recovered a strong radius dependence on metallicity for fully convective companions and that any systematic radius inflation was highly model dependent. The discovery and precise characterisation of more late M dwarfs transiting bright F, G or K dwarf stars in Eclipsing Binary - Low Mass \citep[EBLM;][]{Triaud2013EBLM1,Gomez2014EBLM2, vonBoetticher2017EBLM3, Triaud2017EBLM4, vonBoetticher2019EBLM5, Gill2019EBLM6, Kunovac2020EBLM7, Swayne2021EBLM8, Sebastian2023EBLM9, Duck2023EBLM10, Maxted2023EBLM, Swayne2024EBLM11, Davis2024EBLM12, Sebastian2024EBLM13} systems is vital to understand why the radius inflation problem exists and whether it extends below the convective boundary.

Many of the models proposed rely on magnetic effects to explain the larger than expected radii of M dwarfs \citep{Mullan2001ConvectiveMdwarfs, Chabrier2007mdwarfinflation, LopezMorales2007mdwarfinflation, MacDonald2014Mdwarfactivity, Morales2022mdwarfinflation}. The majority of EBLMs have short orbital periods (over 70\% of the sample from \citet{Maxted2023EBLM} have orbital periods $<$10\,days) and hence their orbital separations are relatively small. In such short-period systems, tidal interactions with the primary are expected to synchronise the rotation of the M dwarf secondary to its orbit, causing it to spin much faster than a typical M-dwarf. This increases the magnetic field strength of the secondary, leading to an inflated radius. However, studies of longer-period M dwarfs have already found conflicting results of how the binary period may affect the radius. For example, \citet{vonBoetticher2019EBLM5} found no relation between orbital period and radius inflation, while \citet{Swayne2024EBLM11} found shorter period binaries were more inflated. However, both of these studies were performed on relatively small samples of M dwarf binary companions with few having periods longer than 30\,days. Thus, to accurately understand radius inflation, a larger sample of longer-period fully convective M dwarf binary companions is required.

It is in this context that we present the discovery of \TICID\,B; a late M dwarf similar in mass and radius to \trappist-1 on a highly eccentric $\sim$193 day orbit with an evolved G type star. Henceforth, we refer to the system as \OBJID\ (see Appendix\,\ref{section:acronym} and Table\,\ref{tab:ngts-eb} for more information on this naming system). In Section\,\ref{section:photobs} we describe the photometric observations that led to the discovery of \OBJID\,B and in Section\,\ref{section:obs:rvs} we describe the spectroscopic observations used to characterise the primary star and to confirm the presence of the secondary companion by radial velocity measurement. In Section\,\ref{section:host-analysis} we describe the techniques used to analyse the spectra in order to characterise the primary and in Section\,\ref{section:orbit_fit} we describe how we fit an orbital solution to the system to obtain parameters for the secondary. In Section\,\ref{section:results} we present the results of our analysis and discuss the nature of the system and its implications for further study of late M dwarfs. Finally, in Section\,\ref{section:conclusion} we summarise the work so far and close with some remarks on potential future work on this system.

\section{Photometric Observations}
\label{section:photobs}

In this section we describe the photometric observations from the Transiting Exoplanet Survey Sattelite \citep[\tess;][Section\,\ref{section:obs:phot:tess}]{ricker2015tess} and the Next Generation Transit Survey \citep[\ngts;][Section\,\ref{section:obs:phot:ngts}]{wheatley2018ngts} that led to the discovery of \OBJID\,B. These observations are summarised in Table\,\ref{tab:phot_summary}.

\begin{table*}
    \centering
    \caption{Summary of the photometric observations for \OBJID.}
    \begin{tabular}{ccccccc}
    \toprule
        Instrument&  Night(s) Observed&  $N_\text{images}$&  Exptime (s)&  Filter& \tess\ Sector&Comments\\
        \hline
        \tess& 2018/12/12-2019/01/06& 973& 1800& \tess& 6&\\
        \tess& 2019/01/08-2019/02/01& 1081& 1800& \tess& 7&\\
        \tess&  2019/02/02-2019/02/27&  807&  1800&  \tess& 8&Transit\\
        \tess& 2020/08/26-2020/09/21& 3572& 600& \tess&29&Transit in ramp\\
        \tess& 2020/12/18-2021/01/13& 3485& 600& \tess& 33&\\
        \tess& 2021/01/14-2021/02/08& 3347& 600& \tess& 34&\\
        \tess& 2021/02/09-2021/03/06& 2711& 600& \tess& 35&\\
        \tess& 2021/05/27-2021/06/24& 3864& 600& \tess& 39&\\
        \tess& 2023/01/18-2023/02/12& 9328& 200& \tess& 61&Quick Look Pipeline only\\
        \tess& 2023/02/12-2023/03/10& 10273& 200& \tess& 62&Quick Look Pipeline only\\
        \tess& 2023/06/02-2023/07/01& 10098& 200& \tess& 66&Quick Look Pipeline only\\
        \ngts& 2022/04/09-2022/06/09& 5088& 10& \ngts& N/A&Initial monitoring\\
        \ngts& 2022/10/19, 2023/02/25, 2023/03/25& 4145& 10& \ngts& N/A&Targeted followup\\
        \ngts& 2023/11/10& 8813& 2& \ngts& N/A&Egress\\
    \bottomrule
    \end{tabular}
    \label{tab:phot_summary}
\end{table*}
\subsection{\tess}
\label{section:obs:phot:tess}

The Transiting Exoplanet Survey Satellite \citep[\tess;][]{ricker2015tess}, is a space-based all-sky survey searching for transiting planets around nearby bright stars. The \tess\ spacecraft has a payload of 4 onboard cameras each consisting of four CCDs. The cameras are arranged in a vertical configuration, resulting in a rectangular field of view. Since 2018, \tess\ has been performing a near all-sky survey by rotating its pointings around the ecliptic poles at regular intervals, with each different pointing referred to as a sector. \tess\ observes each $24\degree \times 96\degree$ sector for $\sim$27\,days at a time, although some targets have longer observation baselines due to the overlapping regions between sectors closer to the ecliptic poles.

\OBJID\ was observed by \tess\ in the Full Frame Images (FFIs) of sectors 6, 7 and 8 at 30-minute cadence; sectors 29, 33, 34, 35 and 39 at 10-minute cadence; and sectors 61, 62 and 66 at 200-second cadence. Sectors 6 to 39 have lightcurves produced from the FFIs by the \tess\ Science Processing Operations Centre \citep[\tess-SPOC;][]{jenkins2016spoc, 2020CaldwellTESSSPOCFFI} pipeline and those for Sectors 61, 62 and 66 are produced by the Quick Look Pipeline \citep[QLP;][]{huang2020qlp}. A summary of the \tess\ lightcurves and other photometric data, including observation dates, is given in Table\,\ref{tab:phot_summary} and the lightcurves are shown in Figure\,\ref{fig:all_tess}. Most of the lightcurves appear relatively featureless with the exception of the transit features in Sector 8 and in the high scattered light region of Sector 29 (this scattered light is removed from the other sectors by detrending). In addition, the Sector 39 lightcurve shows high correlated noise, likely due to spacecraft jitter. Note that we remove any timestamps with flux values of \texttt{NaN} or with a data quality flag not equal to zero (with the exception of sector 29 where a transit occurs near the end of a \tess\ orbit in a region of high scattered light).

A single transit of \OBJID\,B was detected by a custom single-transit search algorithm, as described in \citet{Gill2020monofind}, in the Sector 8 \tess-SPOC FFI lightcurve at TBJD=1519.3. The Sector 8 \tess-SPOC FFI lightcurve containing the transit is shown in Figure~\ref{fig:transits_TESS8}. A second transit occurred in sector 29 but was not visible in the \tess-SPOC lightcurve \texttt{PDCSAP} Flux due to scattered light from the Earth at the end of the orbit. To remove the scattered light, we used the \texttt{RegressionCorrector} class from the \texttt{Lightkurve} \citep{lightkurve2018} package for \texttt{Python}. The \texttt{RegressionCorrector} class uses a `design matrix', made up of the flux from all pixels outside of the aperture with an offset term in order to fit the mean level of the lightcurve. This is done under the assumption that pixels outside the aperture will contain no flux from the target. The lightcurve is then detrended against all of these vectors to produce a model of the mean flux. The resulting model flux is subtracted from itself at the fifth percentile to avoid reducing the flux below zero before being subtracted from the uncorrected lightcurve, resulting in a lightcurve corrected for scattered light. This allowed us to produce a lightcurve from the \tess-SPOC FFI target pixel file (\texttt{tpf}) with the scattered light corrected for in order to reveal the transit event - which is shown in Figure\,\ref{fig:transits_TESS29}.

\subsection{\ngts}
\label{section:obs:phot:ngts}

The Next Generation Transit Survey \citep[\ngts;][]{wheatley2018ngts} is a facility made up of 12 independent robotic telescopes with 20-cm apertures located in Paranal, Chile. The high photometric precision of \ngts, which can rival or even exceed that of \tess\ from the ground \citep{Bryant2020WASP166bNGTS, Bayliss2022NGTSperformance, OBrien2022ScintillationNGTS}, makes it well suited for follow-up of single transit events from \tess, as described in \citet{Bayliss2020NGTSmessenger}.

Following the detection of the initial transit in \tess\ Sector 8 (see Section\,\ref{section:obs:phot:tess}), \OBJID\ was initially monitored by \ngts\ with 10-s exposures on 15 nights from 2022/04/09 to 2022/06/09. After solving the system (see Section\,\ref{section:allesfitter}) a transit was predicted to occur on the night of 2022-04-09. However one was not detected, this is because the observation window only covered the mid transit time with no ingress or egress observed. After the second \tess\ transit was observed, the period could be reduced to a set of discrete aliases, each an integer fraction of the time between the first and second \tess\ transits. \OBJID\ was set up for targeted monitoring with \ngts\ to try and observe possible period aliases on the nights of 2022/10/19, 2023/02/25, 2023/03/25 and 2023/11/10. No transits were seen on 2023/02/25 or 2023/03/25. Similarly to the first set of observations, a mid transit with no baseline was observed on the night of 2022/10/19 - leading to another non-detection. Neither of these nights were included in the final transit fit (see Section\,\ref{section:allesfitter}). Details of this monitoring are set out in \autoref{tab:phot_summary}.

\begin{figure}
    \centering
    \begin{subfigure}{\columnwidth}
        \includegraphics[width=\textwidth]{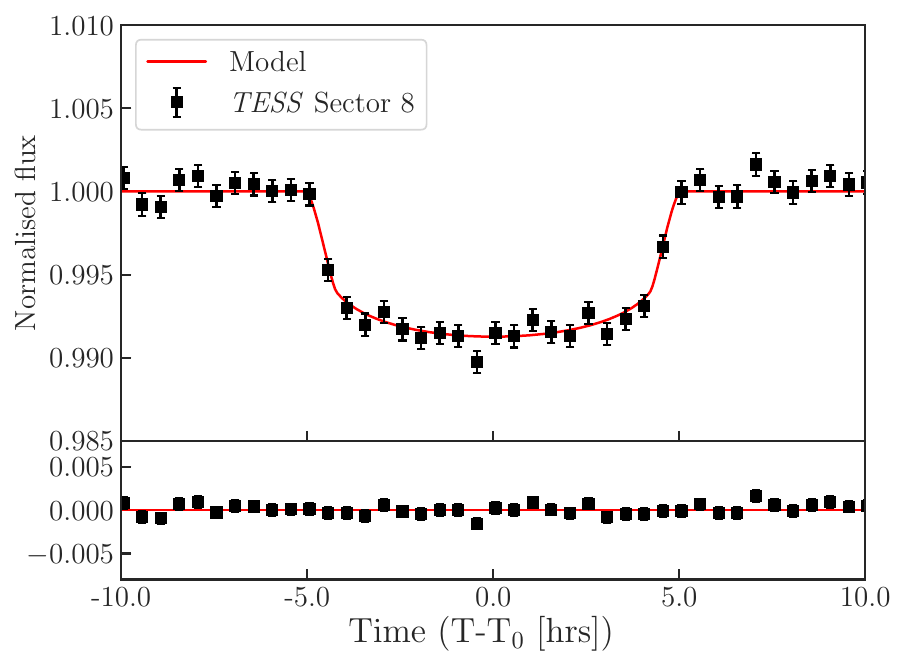}
        \caption{\tess\ Sector 8 transit.}
        \label{fig:transits_TESS8}
    \end{subfigure}
    \begin{subfigure}{\columnwidth}
        \includegraphics[width=\textwidth]{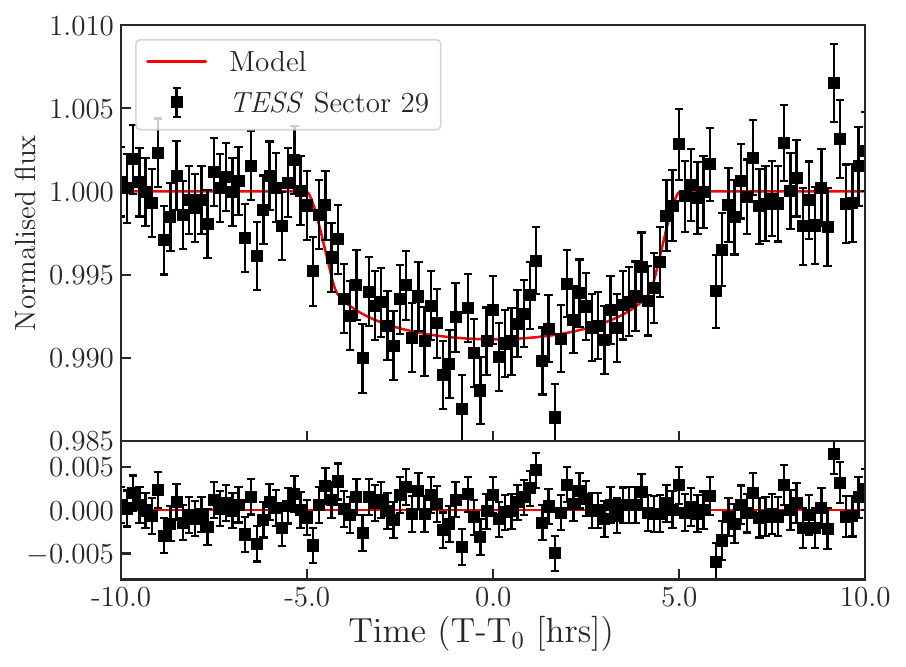}
        \caption{\tess\ Sector 29 transit.}
        \label{fig:transits_TESS29}
    \end{subfigure}
    \begin{subfigure}{\columnwidth}
        \includegraphics[width=\textwidth]{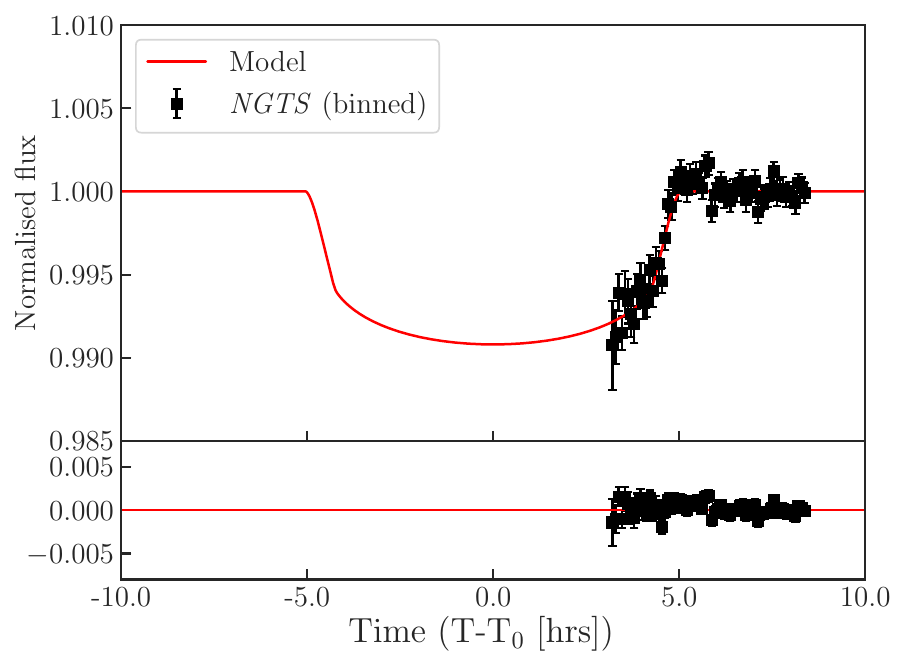}
        \caption{\ngts\ transit.}
        \label{fig:transits_NGTS}
    \end{subfigure}
    \caption{Transit lightcurves of \OBJID\,normalised to the out-of-transit flux levels. Each panel shows the median transit model (see Section\,\ref{section:allesfitter}) as a solid red line. The transit data plotted is shown as black square markers with errorbars. The top panel of each subfigure shows the data and model while the lower panel of each shows the residuals after the model has been subtracted from the observed data.}
    \label{fig:transits}
\end{figure}

A transit egress and baseline of \OBJID\,B was observed for 5.6 hours on the night of 2023/11/10. The event was consistent with the events seen in TESS for shape and depth. The event confirmed the orbital period of the system to be $P\approx193$\,days. We binned the \ngts\ data into 63 bins equally spaced in time with a resulting cadence of 300\,s. The binned lightcurve is shown in Figure\,\ref{fig:transits_NGTS}.

\section{Spectroscopic observations}
\label{section:obs:rvs}

In this section we describe the spectroscopic measurements of \OBJID\ from \coralie\ \citep[][Section\,\ref{section:obs:rvs:coralie}]{queloz2001coralie}, the High Accuracy Radial-velocity Planet Searcher \citep[\harps;][Section\,\ref{section:obs:rvs:harps}]{mayor2003harps} and the Fiberfed Extended Range Optical Spectrograph \citep[\feros;][]{Kaufer1999FEROS}, that were used to characterise the primary star and confirm the presence of \OBJID\,B via radial velocity (RV) measurements. The RV measurements from both instruments can be found in Table\,\ref{tab:rv_data}.

\begin{table}
    \centering
    \caption{Radial velocity measurements of \OBJID.}
    \begin{tabular}{ccccc}
	\toprule
	Time & Radial Velocity & Exposure time & SNR\textbf{*} & Airmass \\
	(BJD-2457000) & $\left(\text{km}\text{s}^{-1}\right)$ & (seconds) & & \\
	\hline
	\multicolumn{5}{c}{\coralie} \\
	\hline
	2546.65717&$79.12\pm0.05$&2100&9.4&1.39\\
	2569.71955&$79.52\pm0.06$&2100&6.4&1.08\\
	2577.71406&$79.67\pm0.03$&2100&11.8&1.08\\
	2583.75381&$79.71\pm0.04$&2100&10.2&1.11\\
	2584.70351&$79.71\pm0.04$&1800&10.2&1.08\\
	2602.63912&$79.76\pm0.02$&2100&15.2&1.08\\
	2608.75240&$79.75\pm0.03$&2101&12.1&1.29\\
	2611.73879&$79.83\pm0.03$&2101&12.0&1.27\\
	2622.72229&$79.77\pm0.03$&2100&13.5&1.31\\
	2630.69521&$79.69\pm0.03$&2100&10.4&1.30\\
	2640.60792&$79.53\pm0.02$&2100&15.3&1.13\\
	2644.57483&$79.54\pm0.04$&2100&9.4&1.10\\
	2648.53875&$79.47\pm0.04$&2100&9.2&1.08\\
	2652.57233&$79.31\pm0.02$&2100&15.1&1.13\\
	2656.57356&$79.14\pm0.03$&2100&13.0&1.15\\
	2680.56890&$74.54\pm0.03$&2100&12.7&1.34\\
	2685.51048&$70.91\pm0.05$&1500&7.6&1.19\\
	2695.51201&$74.64\pm0.06$&1500&5.9&1.28\\
	2700.51825&$76.20\pm0.04$&1500&9.6&1.37\\
	2726.46561&$78.74\pm0.03$&2100&13.5&1.46\\
	3002.63475&$79.81\pm0.05$&1500&8.8&1.17\\
	\hline
	\multicolumn{5}{c}{\harps} \\
	\hline
	2579.82769&$79.546\pm0.005$&1800&25.1&1.28\\
	2582.76821&$79.568\pm0.004$&1800&32.9&1.14\\
	2584.77773&$79.589\pm0.005$&1800&25.9&1.17\\
	2586.77178&$79.593\pm0.004$&1800&27.8&1.17\\
	2601.71653&$79.668\pm0.007$&1800&19.1&1.14\\
	2603.75581&$79.625\pm0.006$&1800&22.3&1.25\\
	2620.64952&$79.662\pm0.004$&1800&28.5&1.12\\
	2621.67762&$79.656\pm0.005$&1800&27.9&1.17\\
	2624.70819&$79.636\pm0.006$&1800&24.3&1.29\\
	2625.73369&$79.631\pm0.008$&1800&19.1&1.42\\
	2649.65720&$79.289\pm0.005$&1800&26.2&1.37\\
	2651.60515&$79.237\pm0.005$&1800&26.3&1.20\\
	2652.64649&$79.186\pm0.009$&1800&16.1&1.35\\
	2653.57857&$79.168\pm0.005$&1800&27.2&1.15\\
    \hline
    \multicolumn{5}{c}{\feros} \\
    \hline
    3268.77318&$70.787\pm0.010$&1200&50&1.10\\
    3266.74627&$70.550\pm0.009$&1200&55&1.19\\
    3267.81058&$70.670\pm0.009$&1200&61&1.09\\
    3268.77318&$71.009\pm0.009$&1200&61&1.13\\
    3311.81894&$78.894\pm0.010$&1200&53&1.24\\
    3313.61205&$79.005\pm0.010$&1200&53&1.21\\
    3366.70585&$79.764\pm0.010$&1200&50&1.39\\
    3368.64831&$79.783\pm0.010$&1200&50&1.20\\
    3381.57640&$79.758\pm0.008$&1200&68&1.12\\
    3408.52850&$79.657\pm0.009$&1200&58&1.17\\
    3437.50366&$78.601\pm0.012$&1200&42&1.33\\
	\bottomrule
\end{tabular}
    \label{tab:rv_data}
    \begin{tablenotes}
    \small
    \item \textbf{*} For \coralie\ we list the recorded SNR of spectral order 62, while for \harps\ we use order 64. The SNR values for \feros\ are described in \S\,\ref{section:ferosobs}.
    \end{tablenotes}
\end{table}

\begin{figure}
    \centering
    \begin{subfigure}{\columnwidth}
        \centering
        \includegraphics[width=\textwidth]{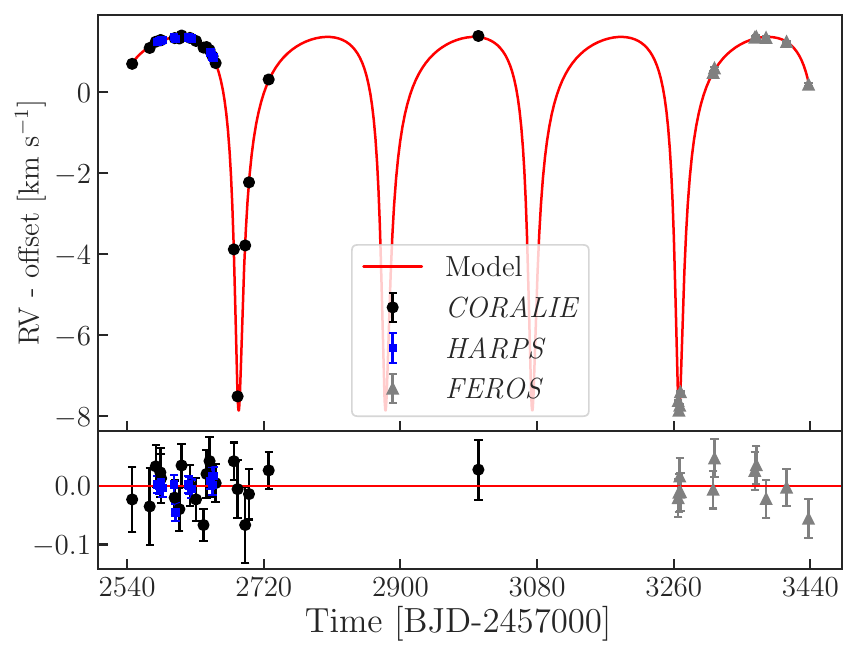}
        \caption{Radial velocities as a function of time.}
        \label{fig:RV_plot_time}
    \end{subfigure}
    \begin{subfigure}{\columnwidth}
        \centering
        \includegraphics[width=\textwidth]{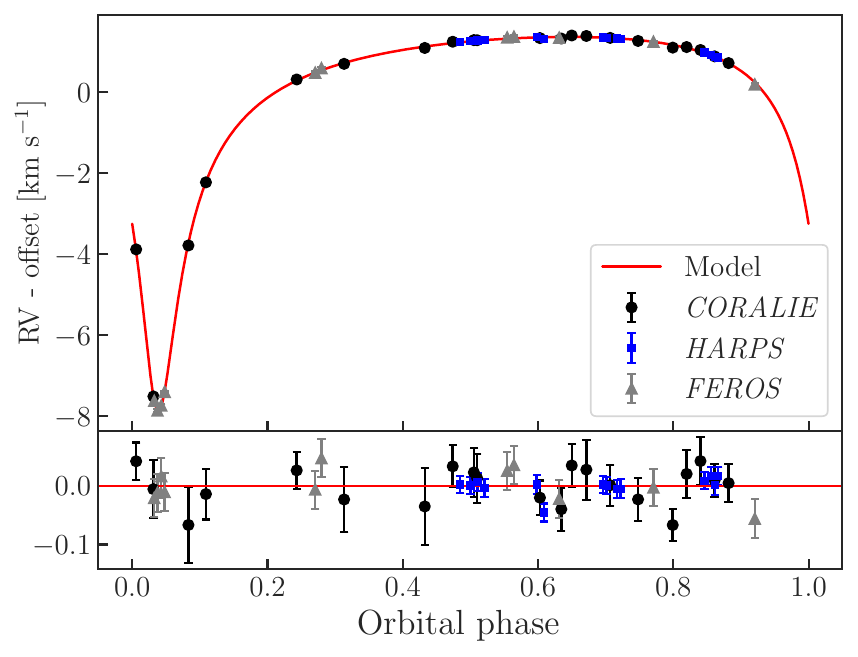}
        \caption{Radial velocities after phase-folding on the orbital period of \Bperiod\,days.}
        \label{fig:RV_plot_phase}
    \end{subfigure}
    \caption[Radial velocity measurements of \OBJID.]{Radial velocity measurements of \OBJID. Data points from \coralie\ are denoted with a black circle, \harps\ is shown with a blue square and \feros\ with grey triangles. The median fitted RV model is also overplotted in red. The top panel shows the data while the bottom shows the residuals after subtracting the best model fit. Subfigure \textbf{(a)} shows the data plotted versus time while \textbf{(b)} shows the same data as a function of orbital phase.}
    \label{fig:RV_plot}
\end{figure}

\subsection{\coralie}
\label{section:obs:rvs:coralie}
We used the \coralie\ \citep{queloz2001coralie} fiber fed echelle spectrograph installed on the 1.2-m Leonhard Euler telescope at ESO La Silla Observatory to take radial velocity measurements of \OBJID\ between the nights of 2021/11/28 and 2023/02/27. A total of 21 spectra were taken within this timeframe. Most of the spectra were taken with an exposure time of 2100\,s with airmasses between 1.46 and 1.08 although an exposure time of 1500\,s was used on 4 nights and 1800\,s on one night. The signal-to-noise ratio (SNR) of order 62 for these spectra ranges between 5.9 and 15.3, leading to an RV precision ranging between 20 and 60\,m\,s$^{-1}$. The data were reduced using the standard \coralie\ data reduction system (DRS 3.8) with a G2 mask used for cross-correlation.  The data are set out in \autoref{tab:rv_data}. These data revealed a large amplitude radial velocity variation in phase with the photometric transit signal. The radial velocity data and best-fitting model (as described in Section\,\ref{section:allesfitter}) are shown in Figure\,\ref{fig:RV_plot}, and indicates that the eclipsing body is a low-mass star on a highly eccentric orbit.

\subsection{\harps}
\label{section:obs:rvs:harps}
We also used the High Accuracy Radial velocity Planet Searcher \citep[\harps;][]{mayor2003harps} spectrograph mounted on the 3.6-m telescope at ESO La Silla Observatory to make additional radial velocity measurements of \OBJID. Between the nights of 2021/12/31 and 2022/03/15, 14 spectra were taken of \OBJID\ using \harps \footnote{Based on observations collected at the European Southern Observatory under ESO programme 108.22L8.001}. All of these observations used an exposure time of 1800\,s, and the airmass ranged from 1.42 to 1.14. The spectra were reduced using DRS 3.0.0 and a G8 mask. The resultant RV measurements are presented in \autoref{tab:rv_data} and shown in Figure\,\ref{fig:RV_plot}.

\subsection{\feros}
\label{section:ferosobs}
We also made use of the Fiberfed Extended Range Optical Spectrograph \citep[\feros;][]{Kaufer1999FEROS} mounted on the 2.2-m MPG/ESO telescope at ESO La Silla to take further RV measurements of \OBJID. These observations were performed in the context of the Warm gIaNts the tEss (WINE) collaboration \citep{brahm:2019,schlecker:2020,hobson:2021,trifonov:2023,brahm:2023,jones:2024}. Between the nights of 2023/11/17 and 2024/05/07, 11 spectra were taken with exposure times of 1200 s at airmasses ranging between 1.09 and 1.25 resulting in SNRs ranging between 42 and 68. Note that these SNR values are per resolution element and are based on the order centered on the Magneisum triplet around 5150\AA. The spectra were reduced using the \texttt{ceres} pipeline \citep{ceres}, resulting in the RV measurements shown in Table\,\ref{tab:rv_data} and Figure\,\ref{fig:RV_plot}. The RV points taken with \feros\ include 4 points clustered around the time of periastron, which proved especially useful in constraining the shape of the RV curve (see Section\,\ref{section:allesfitter}).

\section{Primary Stellar Analysis}
\label{section:host-analysis}

\subsection{Catalogue parameters}
\label{section:catalog-params}
Catalogue parameters for \OBJID\ are shown in Table~\ref{tab:host_properties}. These were taken from the 2 Micron All Sky Survey \citep[2MASS;][]{skrutskie20062mass}, \gaia\ data release 3 \citep[\gaia\ DR3;][]{gaia2021dr3}, the \tess\ Input Catalogue v8 \citep[TIC 8;][]{Stassun2019tic}, the AAVSO Photometric All Sky Survey \citep[APASS;][]{Henden2014APASS}, the Wide field Infrared Survey Explorer \citep[WISE;][]{wright2010wise} and SkyMapper \citep{Keller2007SkyMapper} catalogues where specified.
\begin{table}
    \centering
    \caption{\OBJID\,A stellar parameters.}
    \begin{tabular}{ccc}
        \toprule
        Property&  Value & Source\\ \hline
        \hline
        \multicolumn{3}{c}{\textit{Identifiers}} \\ 
        \hline
        2MASS ID&  J06564704-5204263& 2MASS (\S\ref{section:catalog-params}) \\
        \gaia\ Source ID&  5504617415848984320& \gaia\ DR3 (\S\ref{section:catalog-params}) \\
        TIC ID&  238060327& TIC 8 (\S\ref{section:catalog-params}) \\
        \hline
        \multicolumn{3}{c}{\textit{Coordinates}} \\ 
        \hline
        RA&  $6^\text{h}56^\text{m}47.04^\text{s}$& \gaia\ DR3 (\S\ref{section:catalog-params}) \\
        DEC&  $-52\degree04{'}26.11^{''}$& \gaia\ DR3 (\S\ref{section:catalog-params}) \\
        \hline
        \multicolumn{3}{c}{\textit{Proper motion and parallax}} \\ 
        \hline
        $\mu_\text{RA}$ (mas y$^{-1}$)&  $4.279\pm 0.013$& \gaia\ DR3 (\S\ref{section:catalog-params}) \\
        $\mu_\text{DEC}$ (mas y$^{-1}$)&  $16.980\pm0.013$& \gaia\ DR3 (\S\ref{section:catalog-params}) \\
        Parallax (mas)&  $2.21\pm0.01$& \gaia\ DR3 (\S\ref{section:catalog-params}) \\
        Radial velocity (\kms) & $78.376\pm0.005$ & This work (\S\ref{section:diskmember}) \\
        \hline
        \multicolumn{3}{c}{\textit{Galactic Kinematics}} \\ 
        \hline
        $U$ (\kms)& $-33.05\pm0.16$& This work (\S\ref{section:diskmember}) \\
        $V$ (\kms)&$-62.86\pm0.02$& This work (\S\ref{section:diskmember}) \\
        $W$ (\kms)&$-0.84\pm0.09$& This work (\S\ref{section:diskmember}) \\
        $v_\text{tot}$ (\kms)& $71.02\pm0.08$& This work (\S\ref{section:diskmember}) \\
        $e_\text{gal}$&0.31& This work (\S\ref{section:diskmember}) \\
        $J_z$ (kpc \kms)&1278& This work (\S\ref{section:diskmember}) \\
        \hline
        \multicolumn{3}{c}{\textit{Magnitudes}} \\ 
        \hline
        V (mag)&  $12.440\pm 0.026$& APASS (\S\ref{section:catalog-params}) \\
        B (mag)& $13.176\pm 0.024$&APASS (\S\ref{section:catalog-params}) \\
        u (mag)& $14.371\pm0.024$&SkyMapper (\S\ref{section:catalog-params}) \\
        g (mag)& $12.596\pm0.019$&SkyMapper (\S\ref{section:catalog-params}) \\
        r (mag)& $12.257\pm0.030$&SkyMapper (\S\ref{section:catalog-params}) \\
        i (mag)& $12.083\pm0.010$&SkyMapper (\S\ref{section:catalog-params}) \\
        z (mag)& $12.096\pm0.020$&SkyMapper (\S\ref{section:catalog-params}) \\
        G (mag)& $12.2787\pm0.0002$&\gaia\ DR3 (\S\ref{section:catalog-params}) \\
        BP (mag) & $12.6455\pm 0.0006$&\gaia\ DR3 (\S\ref{section:catalog-params}) \\
        RP (mag) &$11.7458\pm 0.0004$&\gaia\ DR3 (\S\ref{section:catalog-params}) \\
        \tess\ (mag)& $11.806\pm0.006$&TIC 8 (\S\ref{section:catalog-params}) \\
        J (mag)& $11.172\pm 0.024 $&2MASS (\S\ref{section:catalog-params}) \\
        H (mag)& $10.858\pm 0.025$&2MASS (\S\ref{section:catalog-params}) \\
        K (mag)& $10.774\pm0.024$&2MASS (\S\ref{section:catalog-params}) \\
        W1 (mag)& $10.715\pm 0.022$&WISE (\S\ref{section:catalog-params}) \\
        W2 (mag)& $10.765\pm 0.020$&WISE (\S\ref{section:catalog-params}) \\
        \hline
        \multicolumn{3}{c}{\textit{Fitted Atmospheric Parameters}}\\
        \hline
        \teff\ (K) & \hostteff & \paws\ (\S\ref{section:PAWS}) \\
        \logg\ ($\log(\text{cgs})$) & \hostlogg & \paws\ (\S\ref{section:PAWS}) \\
        \met\ (`dex') & \hostmet & \paws\ (\S\ref{section:PAWS}) \\
        \logrhk & \logrhkval & This work (\S\ref{section:activity}) \\
        \vsini\ (\kms) & \hostvsini & \paws\ (\S\ref{section:PAWS})\\
        \vmic\ (\kms) & $1.42\pm0.03$ & \paws\ (\S\ref{section:PAWS}) \\
        \vmac\ (\kms) & $3.92$ & \paws\ (\S\ref{section:PAWS}) \\
        \hline
        \multicolumn{3}{c}{\textit{Derived parameters}} \\ 
        \hline
        \rstar\ (\rsun) & \hostrad & \isochrones\ (\S\ref{section:isochrones}) \\
        \noalign{\smallskip}
        \mstar\ (\msun) & \isohostmass & \isochrones\ (\S\ref{section:isochrones}) \\
        \noalign{\smallskip}
        Age (Gyr) & \hostage & \isochrones\ (\S\ref{section:isochrones}) \\
        \bottomrule
    \end{tabular}
    \label{tab:host_properties}
\end{table}

\subsection{Spectral fitting}
\label{section:PAWS}
To fit accurate stellar atmospheric parameters for the primary we combined the 14 individual \harps\ spectra taken for \OBJID\ into a single spectrum with an SNR value of 85. In addition, we masked out certain regions of the spectra that were affected by systematics due to ghosting on the CCD, these masked regions are listed in Table\,\ref{tab:bumps}.

We used the \paws\ pipeline to analyse the \harps\ spectra for \OBJID\ in order to obtain accurate stellar parameters for the primary star. It is described in full in \citet{Freckelton2024paws}, however we summarise its function in the remainder of this section. \paws\ is built on the \texttt{ispec} package \citep{Blanco-Cuaresma2014ispec, Blanco-Cuaresma2014ispeccode}. It uses the line list created for the \texttt{spectrum} package \citep{Gray-Corbally1994SPECTRUM} with some minor additions described in \citet{Freckelton2024paws}, as well as the ATLAS 9 set of model atmospheres \citep{Kurucz2005ATLAS} to perform fits to spectra using a combined equivalent widths and synthesis method. 

The equivalent widths fitting is done on only the Fe\,{\sc I} and Fe\,{\sc II} lines utilizing the \texttt{ARES} \citep{Sousa2015ares} and \texttt{WIDTH} \citep{Sbordone2004Width} software packages and the solar abundances collated by \citet{Blanco-Cuaresma2019solarabundance}. The effective temperature (\teff), surface gravity (\logg)\ and metalicity (\met)\ are estimated using this method and, in turn, are then used to estimate a value of the microturbulent velocity (\vmic)\ by applying the \gaia-ESO (GES) relation based on the \gaia-ESO survey UVES data release 1 as described in \citet{Blanco-Cuaresma2014ispec} and \citet{Jofre2014GES}. 

The values of \teff\ and \logg\ derived from equivalent widths are then used as priors for the synthesis fitting method while the values of \met\ and \vmic\ are fixed. The \texttt{SPECTRUM} package \citep{Gray-Corbally1994SPECTRUM} is then used to generate synthetic spectra that are iteratively fit to the observed data. Values of \teff, \logg\ and the projected stellar rotational velocity (\vsini)\ are fit for and the macroturbulent velocity (\vmac)\ is calculated using the same GES derived relation used for \vmic\ in the portion of the pipeline that determines the equivalent widths. Six iterations were used, which was determined to be the optimal number by \citet{Blanco-Cuaresma2014ispec}. As is standard for \paws, a value of 100\,K is added in quadrature to the errors in \teff\ as otherwise the errors reported by \texttt{ispec} are more precise than the expected accuracy of the atmospheric models used \citep[see][for details]{Freckelton2024paws}.

The fitted parameters are shown in Table\,\ref{tab:host_properties}. These values show a temperature consistent with an early G-type star and enriched metalicity as well as a surface gravity consistent with a star evolving off the main sequence. These values were used as priors in order to determine a mass and radius for the star via isochrone model fitting (see Section\,\ref{section:isochrones}).

\subsection{Isochrone fitting}
\label{section:isochrones}
We used the \isochrones\ package \citep{Morton2015isochrones} to determine a mass and radius of the primary by fitting MIST isochrones \citep{dotter2016isochrones, Choi2016isochrones}. Following \citet{Davis2024EBLM12} we used the J, H and K infrared band catalogue photometry points as well as the \gaia\ parallax shown in Table\,\ref{tab:host_properties} and the \paws\ fitted parameters as priors. The resulting parameters are shown in Table\,\ref{tab:host_properties}. The uncertainties were estimated by extracting the 84th and 16th percentiles of the posterior sample. To reflect the inherent systematic uncertainty in evolutionary models, we follow the recommendations of \citet{Tayar2022isochroneerrors} and add an additional 4.2\% error term in quadrature to the stellar radius. We also add errors in quadrature to the mass, and age values of 0.03\,\msun and 1\,Gyr, respectively, found by taking the standard deviation of fitted values from different sets of isochrones using the \texttt{kiauhoku} package \citep{Tayar2022isochroneerrors}.

Since we already knew the star was in a binary, we fitted both a double and single star model but found the parameters for the secondary were in strong disagreement with those obtained from our global fit of the system (see Section~\ref{section:allesfitter}). This is likely due to the secondary contributing a very small amount of light to the passbands used, meaning \isochrones\ cannot obtain accurate secondary stellar parameters from these wide band photometric points. The primary parameters changed very little between these two fits, hence we choose to present the single star fit for simplicity.

\subsection{Checking against other sources}
\label{section:checking_spectral_fits}
In order to provide a check on the parameters derived using \paws\ and \isochrones\ we compare the values obtained from these methods to those from other sources. We use the \gaia\ DR3 and TICv8 catalogues. We also use values obtained from fitting to the spectral energy density (SED) using the \ariadne\ package \citep{vines2022ariadne}. Another set was obtained by matching the target spectrum against a library of spectra using \specmatch\ \citep{yee2017specmatch}. We also obtained an additional parameter set by fitting against the spectrum using \texttt{species} \citep{SotoandJenkins2018Species}. We find reasonable agreement across all these sources apart from the \gaia\ metallicity, which is $\sim-0.2$\,`dex'. However, due to the relatively small region of the spectrum sampled by \gaia, this is likely a systematic effect. This leads us to conclude the values from \paws\ and \isochrones\ are accurate and we adopt these values for the primary going forward.

\subsection{Stellar activity and rotation}
\label{section:activity}
A visual inspection quickly shows the out-of-transit \tess\ and \ngts\ lightcurves (see Section\,\ref{section:photobs}) to be flat and a Lomb-Scargle periodogram created using \texttt{lightkurve} \citep{lightkurve2018} also returned no significant peaks.

We also calculated \logrhk\, \citep[following the procedure set out in][]{Lovis2011HARPSlogrhk,Costes2021activity} using the coadded \harps\ spectrum as the individual spectra had insuffcient SNR. First we measured the flux at two pass bands centred on the Ca\,{\sc II} H and K line cores. The $H$ band is centred at 3968.47\,\AA\ whilst the $K$ band is centred at 3933.664\,\AA\ and both have a triangular shape and a Full Width Half Maximum (FWHM) of 1.09\,\AA. We then measured the flux of two continuum passbands with widths of 20\,\AA\ centered at 4001.070\,\AA\ $(R)$ and 3901.070\,\AA\ $(V)$. These passband fluxes were then used to calculate the Mt Wilson $S$ index using the following equation:

\begin{equation}
    S = 1.111 \times \frac{H+K}{R+V} + 0.0153.
    \label{eq:S_index}
\end{equation}

\noindent The multiplication coefficient of 1.111 and additive constant of 0.0153 are calibration terms from \citet{Lovis2011HARPSlogrhk} that map the flux ratios calculated from \harps\ spectra to the Mt Wilson $S$ index. We then use this value to calculate R$^\prime_\mathrm{HK}$ as described in \citet{Noyes1984RprimeHandK}:

\begin{equation}
    \text{R}^\prime_\mathrm{HK} = 1.340\times10^{-4} \times C_\text{cf}(B-V) \times S \times R_\text{phot}(B-V),
    \label{eq:RprimeHandK}
\end{equation}

\noindent where $C_\text{cf}(B-V)$ and $R_\text{phot}(B-V)$ are both correction factors based on the photometric $B-V$ colour calculated from magnitudes shown in Table\,\ref{tab:host_properties} that account for varying flux in continuum passbands and the photospheric contribution in the $H$ and $K$ passbands, respectively. We find a value of \logrhk=\logrhkval.

In addition, we used equation 3 and 4 from \citet{Noyes1984RprimeHandK} to estimate the rotation period of \OBJID\,B from \logrhk\ and B$-$V, finding a value of $39\pm6$\,d. When we compare this to the upper limit on the rotational period as determined from the stellar radius (\rone) and projected equatorial rotation velocity (\vsini) using equation 1 from \citet{Watson2010estimatingplanetmass} of P$_\text{rot}\sin i_\star = 28\pm8$\,d, we find these values are in reasonable agreement. However, we note that this relation between \logrhk\ and rotational period \citep[and others e.g;][]{Mamajek2008Agerotlogrhk, Stanford-Moore2020Baffles} is calibrated for main sequence stars and may not give an accurate rotation rate for a star evolving off the main sequence. For a simple sanity check on \logrhk, however, the \citet{Noyes1984RprimeHandK} relation is adequate.

In conclusion, a \logrhk=$-5.12\pm0.17$ strongly suggests that \OBJID\,A is a low-activity star. This, along with its relatively slow rotation period, provides additional evidence towards its nature as an old subgiant evolving off the main sequence since older stars are expected to spin-down via magnetic breaking \citep{Kraft1967stellarrot, Kraft1970stellarrot}, reducing their rotation rate and level of magnetic activity \citep[see figure 6 in][]{Jenkins2011Subgiantlogrhk}.

\section{Orbital solution}
\label{section:orbit_fit}
\subsection{Global modelling with \alles}
\label{section:allesfitter}
We used the \alles\ \citep{allesfitter-paper, allesfitter-code} package for \texttt{Python} to simultaneously fit the photometric transit data and spectroscopic radial velocity data for \OBJID\ (see \autoref{section:photobs} and \autoref{section:obs:rvs}). \alles\ combines other \texttt{Python} packages including \texttt{ellc} \citep{maxted2016ellc} for modelling light-curves, \texttt{emcee} \citep{foreman-mackey2013emcee} for Monte-Carlo Markov-Chain (MCMC) sampling, \texttt{dynesty} \citep{speagle2020dynesty} for nested sampling \citep{Skilling2004nestedsampling, Skilling2006nestedsampling} and \texttt{celerite} \citep{foreman-mackey2017celerite} for Gaussian processes, among others.

We used a nested sampling approach \citep{Skilling2004nestedsampling, Skilling2006nestedsampling} to fit a global model to \OBJID. The photometric solution for the orbital period was $\approx193$\,days and the RV periodogram also returned a peak with a false alarm probability $<1\%$ at this value. Hence, we set a uniform prior on the orbital period $(P)$ between 192 and 194\,days. Additionally, a fit was carried out on just the \tess\ Sector 8 data to obtain priors on the epoch ($T_{0;\text{B}}$), radius ratio ($R_\text{B} / R_\text{A}$), semi-major axis ($(R_\text{A} + R_\text{B}) / a$) and orbital inclination ($\cos{i_B}$) of the system. The eccentricity and argument of periastron of the system were parameterised as the terms $\sqrt{e} \cos{\omega}$ and $\sqrt{e} \sin{\omega}$ and fitted each with uniform priors between $-1$ and 1. Similarly we fit the inclination of the system ($\cos i$) using a uniform prior between 0 and 1. We used a quadratic limb darkening model described in \cite{kipping2013limbdarkening} and used the Limb Darkening ToolKit \citep[\texttt{ldtk};][]{parviainen2015ldtk} package to determine priors on the limb darkening coefficients ($q_1$ and $q_2$, for each instrument) based on the observing filters and primary star properties.

To fit the out-of-transit variability for the \tess\ lightcurves (see Section~\ref{section:obs:phot:tess}) we used Gaussian processes (GPs) with a Matern 3/2 kernel. There was not sufficient out-of-transit data available for the night of the \ngts\ egress observation to use a GP so we did not model out-of transit variation for the \ngts\ data (see Section~\ref{section:obs:phot:ngts}), however we found the variability out of transit for these data was sufficiently low to justify this approach. However, we did include an additional error term ($\log \sigma$) for \ngts\ assuming white noise. Priors on the GP terms ($\ln \sigma $ and $\ln \rho$) and the additional error term ($\log \sigma$) for each \tess\ sector were obtained by using the \texttt{estimate\_noise\_out\_of\_transit} function included in the \alles\ package to perform an MCMC fit with 100 walkers, 2000 burn steps and 3000 total steps to fit these parameters to the out of transit data. This same method was also used to obtain a prior on the error term for \ngts. In addition, we included jitter terms $(\log \sigma)$ for \harps, \coralie\ and \feros\ in the fit. We also fitted for a radial velocity baseline/offset term for each instrument using the \gaia\ DR3 measured radial velocity of the system as a prior (see Table~\ref{tab:host_properties}). The combined RV data plotted in Figure\,\ref{fig:RV_plot} was produced by subtracting the fitted baseline term for each instrument (which are reported in Table~\ref{tab:ns_table}) from the data.. For the semi-amplitude ($K$) we set a uniform prior between 0 and 12 \kms.

It should be noted that we do not include a surface brightness ratio parameter in our global model fit due to the considerably lower flux contribution from the secondary. Using a maximum estimate of the flux ratio in the \tess\ band (see Section\,\ref{section:occ_depth}) we find that the contribution of the secondary  to the total binary flux is less than 0.05\%. The dilution effect from the secondary flux is thus only able to affect our derived secondary radius by at most $\sim0.22\%$, which is over an order of magnitude smaller than the statistical uncertainty on the radius that we ultimately obtain (see Table~\ref{tab:secondary_params}).

The transit and RV data is shown with the fitted models in Figures\,\ref{fig:transits}\,and\,\ref{fig:RV_plot}, respectively, while a selection of parameters fitted from the global model can be found in \autoref{tab:secondary_params}. The full set of parameters outputted by \alles\ can be found in Appendix\,\ref{section:allesfitter_tables} with fitted parameters and their priors shown in Table\,\ref{tab:ns_table} and derived parameters in Tables\,\ref{tab:ns_derived_table1}\,and\,\ref{tab:ns_derived_table2}. In addition the default \alles\  corner plots are shown in Appendix\,\ref{section:allesfitter-plots}.

\subsection{Modified Mass function}
\label{section:mass_function}
We follow the method described in \citet{Davis2024EBLM12} to semi-empirically determine the masses of the primary and secondary, with only the radius of the primary (\rone) being model dependent and every other parameter being either a constant or an observable we can fit for as described in Section\,\ref{section:allesfitter}. We perform all of the calculations in this section using the full posterior sample from \alles\ for the observables and from \isochrones\ for \rone. The observables used in these equations are not directly fitted by \alles\ except for the orbital period, $P$, but can be calculated entirely using allesfitter observables as shown below:

\begin{equation}
    \frac{a}{\rone} = \frac{1+\left(\rtwo/\rone\right)}{\left(\rone+\rtwo\right)/a}
    \label{eq:R1overa_from_fit}
\end{equation}

\begin{equation}
    e = \left(\sqrt{e}\sin\omega\right)^2+\left(\sqrt{e}\cos\omega\right)^2
    \label{eq:ecc_from_fit}
\end{equation}

\begin{equation}
    \sin i = \sqrt{1-(\cos i)^2}
    \label{eq:sini_from_fit}
\end{equation}

To begin deriving the parameters, we take the primary star density as defined by Kepler's law:

\begin{equation}
    \frac{\mone}{\rone^3} = \frac{4\pi^2}{G\,P^2}\left(\frac{a}{\rone}\right)^3 - \frac{\mtwo}{\rone^3},
    \label{eq:host_density}
\end{equation}

\noindent where $G$ is the gravitational constant, $a$ is the orbital semi-major axis, $P$ the orbital period and \mone, \mtwo, \rone\ and \rtwo\,are the masses and radii of the primary and secondary, respectively.
We can then rearrange Equation\,\ref{eq:host_density} for the total mass of the binary:

\begin{equation}
    M = \mone + \mtwo = \frac{4\pi^2}{G\,P^2}\left(\frac{a}{\rone}\right)^3 \rone^3.
    \label{eq:total_mass}
\end{equation}

\noindent For \OBJID\,AB, we find a total mass of $M=1.14^{+0.06}_{-0.08}$\,\msun. We then calculate the modified mass function $f_m$. We refer to this mass function as `modified' since it contains a $\sin^3i$ term, which is absent from the typical spectroscopic mass function since the inclination is nearly impossible to determine from spectroscopic observations alone. However, since \OBJID\,AB is eclipsing, we can include this term as shown in the equation taken from \cite{Hilditch2001closebinarystars} below:

\begin{equation}
    f_m = \frac{\left(1-e^3\right)^{3/2}}{\sin^3i}\frac{PK^3}{2\pi G}.
    \label{eq:mass_function}
\end{equation}

\noindent From this, we determine a value of $f_m=0.0000675\pm0.000005$\,\msun. This mass function can also be expressed as:

\begin{equation}
    f_m = \frac{\mtwo^3}{\left(\mone+\mtwo\right)^2},
\end{equation}

\noindent which can be substituted into Equation\,\ref{eq:total_mass} and solved for the primary and secondary mass:

\begin{equation}
    \mtwo = \sqrt[3]{f_m M^2},
\end{equation}

\noindent for which we obtain a value of $\mtwo=0.096\pm0.002$\,\msun and $\mone=M-\mtwo=1.05^{+0.06}_{-0.07}$\,\msun. Our value of \mtwo\ obtained using this method is smaller than that obtained from \isochrones\ by 0.08\,\msun. However, the difference between both values is only slightly $>1\sigma$ and thus is unlikely to be significant.

\section{Results and discussion}
\label{section:results}

\subsection{The \OBJID\,AB system}
\label{section:overview}

\begin{figure}
    \centering
    \includegraphics[width=\columnwidth]{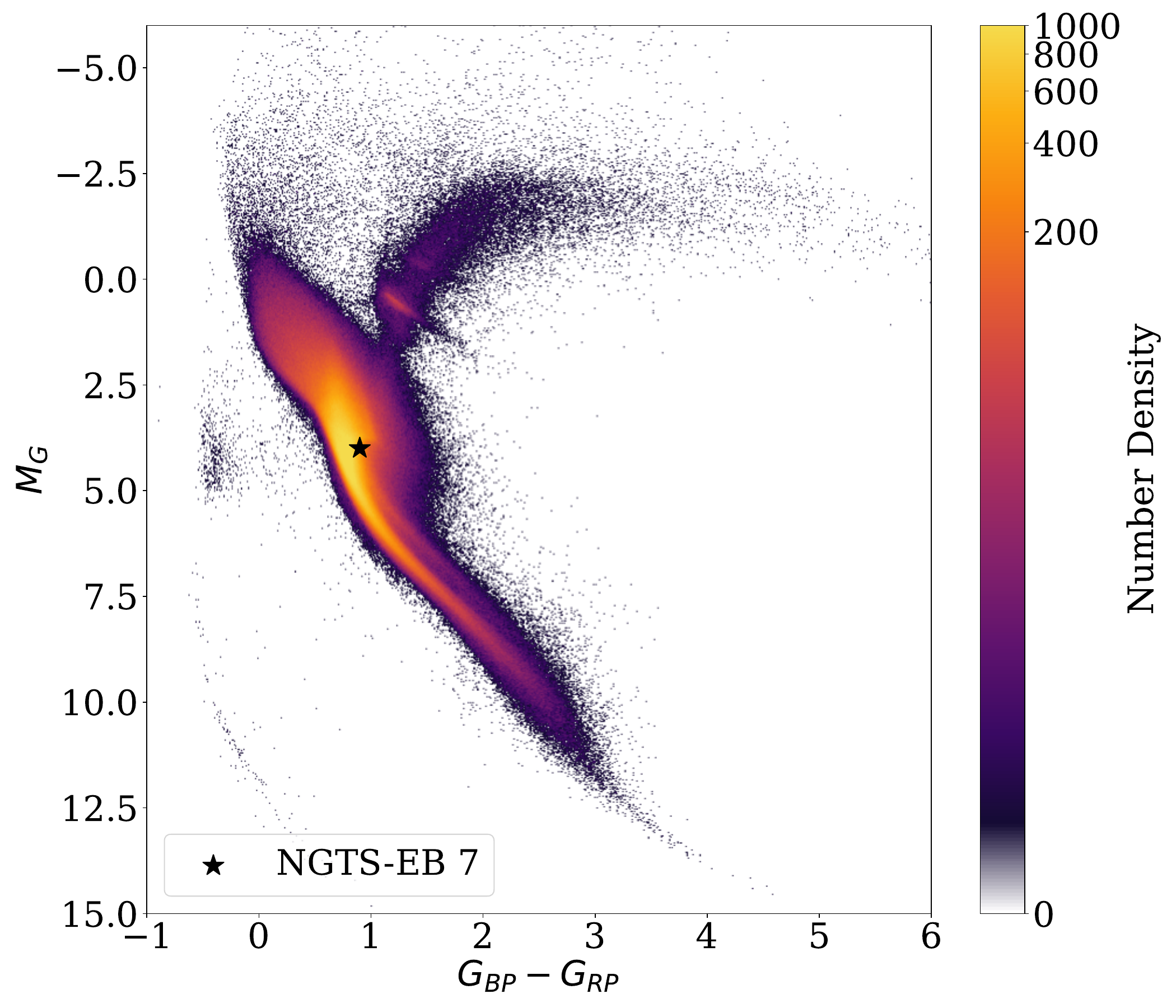}
    \caption{\gaia\ DR3 Hertzsprung Russell diagram with the position of \OBJID\,A highlighted with a black star symbol. \gaia\ BP-RP colour is plotted against absolute magnitude in the G band for all stars in the crossmatched SPOC FFI sample described in \protect\citet{Doyle2024gaiaSPOC}. The position of \OBJID\,A shown is consistent with a star beginning to move off the main sequence.}
    \label{fig:gaia_HRD}
\end{figure}

We find \OBJID\,A is likely a metal rich (\met=\hostmet `dex') evolved early G-type star, with an effective temperature of \hostteff\,K. The enhanced metalicity of the primary is somewhat unusual and falls towards the upper end of what is expected for binary systems \citep[see figure 9 in;][]{Jenkins2015binarymetalicity}. The evolved nature of the primary is apparent from the low value of \logg=\hostlogg\ that we measured from its spectra as well as its position on a Hertzsprung Russell (HR) diagram (see Figure\,\ref{fig:gaia_HRD}). We find a radius of \hostrad\,\rsun\ and a mass of \isohostmass\,\msun. The evolved nature of the primary also allows us to strongly constrain its age to \hostage\,Gyrs.

\begin{figure*}
    \begin{subfigure}{\textwidth}
        \centering
        \includegraphics[width=\textwidth]{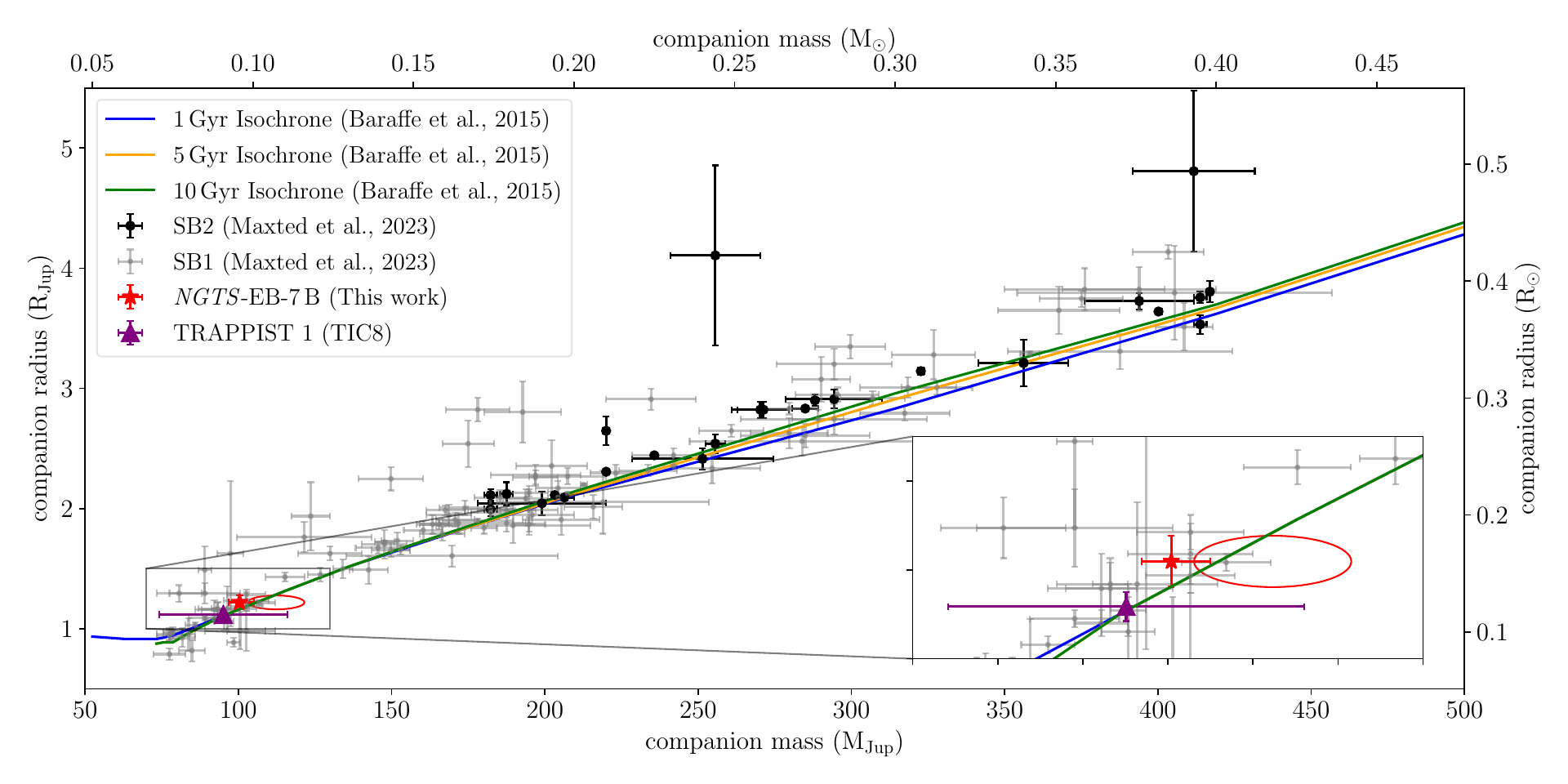}
        \caption{Mass v Radius}
        \label{fig:mass-radius}
    \end{subfigure}
        \begin{subfigure}{\columnwidth}
        \centering
        \includegraphics[width=\textwidth]{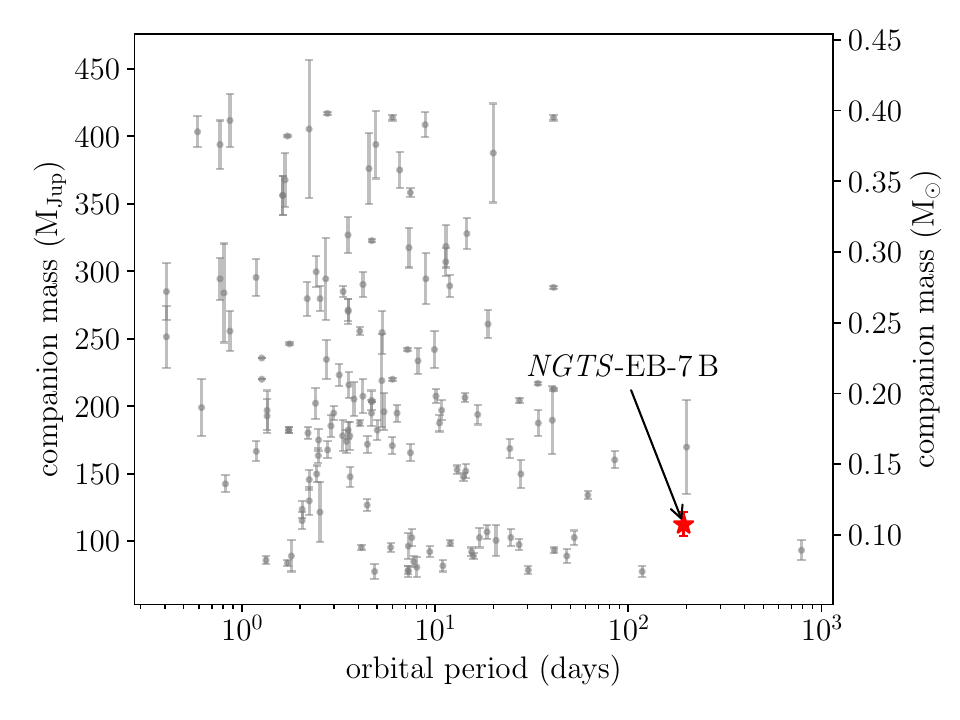}
        \caption{Period v mass}
        \label{fig:period-mass}
    \end{subfigure}
    \begin{subfigure}{\columnwidth}
        \centering
        \includegraphics[width=\textwidth]{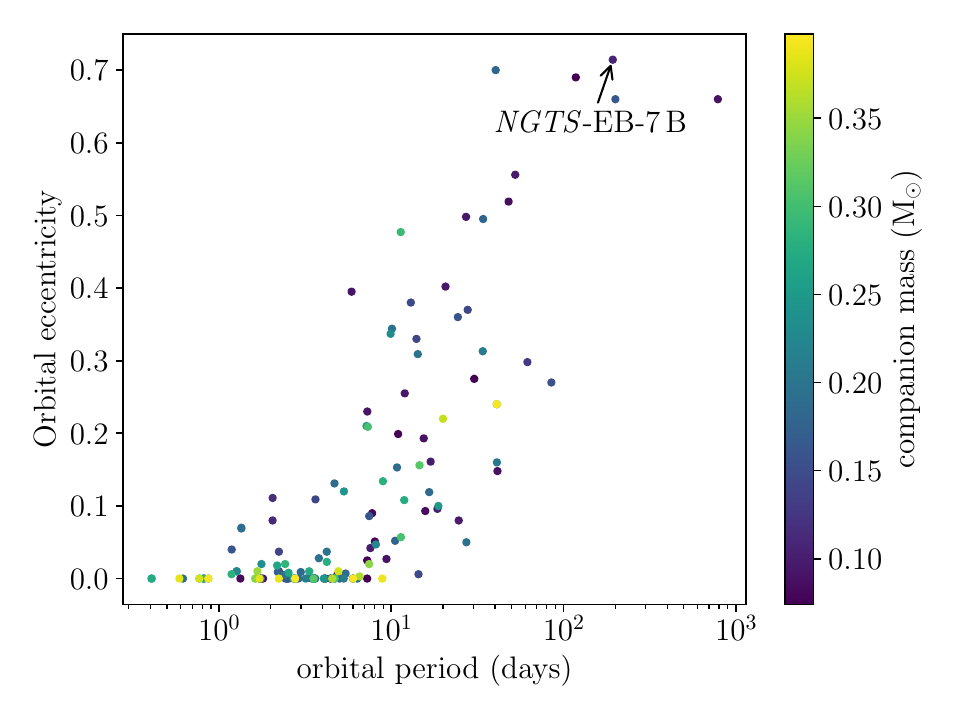}
        \caption{Period v eccentricity}
        \label{fig:period-eccentricity}
    \end{subfigure}
    \caption{Plots of the low-mass companions of eclipsing binaries as taken from the catalogue of \protect\citet{Maxted2023EBLM} which contains all systems from the EBLM project and other systems with a quoted precision of 5\% or less. \OBJID\,B is shown as a red star for comparison. \textbf{Top (a):} Mass radius plot of EBLM companions. Single lined binaries (SB1) are shown by grey dot markers while double lined binaries (SB2) are shown with black circle markers. In addition to the red star with errorbars representing the mass and radius of \OBJID\,B as calculated in Section\,\protect\ref{section:mass_function}, the uncorrected values from \alles\ are shown with a red ellipse (note that both are within 1-$\sigma$ agreement -- see Section\,\protect\ref{section:secondary_inflation}). \trappist-1 is also plotted as a purple triangle for comparison. Also shown is the 1, 5 and 10 Gyr isochrones from \protect\citet{Baraffe2015Mdwarfmodels} plotted as blue, orange and green solid lines, respectively. An inset axis is shown zoomed in around \OBJID\,B. \textbf{Bottom left (b):} Orbital period plotted on a logarithmic scale against companion mass. \textbf{Bottom right (c):} Orbital period plotted on a logarithmic scale against orbital eccentricity and coloured by mass. Note that the error in eccentricity is too small to be seen.} 
    \label{fig:EBLM-pops}
\end{figure*}

For the secondary stellar companion, our orbital solution from fitting a global model (see Section~\ref{section:allesfitter}) finds a mass of \MBMsun\,\msun\ (\MBMjup\,\mjup) and a radius of \BRcompanionRsun,\rsun. As can be seen in Figure\,\ref{fig:mass-radius}, this places \OBJID\,B as a late M dwarf, similar in size to \trappist-1 \citep{Gillon2017TRAPPIST1} and close to the hydrogen burning limit \citep[$\approx$80\mjup;][]{Chabrier2023hydrogenlimit} and the boundary between main sequence stars and brown dwarfs. 

\begin{figure}
    \centering
    \includegraphics[width=\columnwidth]{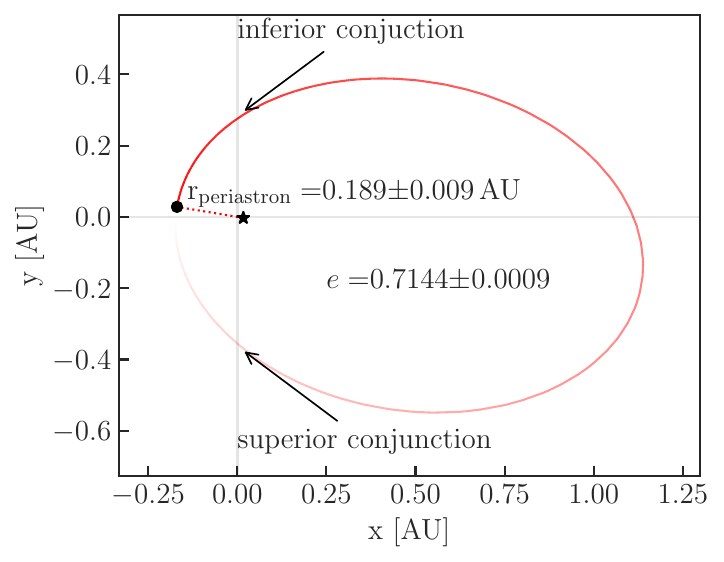}
    \caption{Face down orbital view of the \OBJID\,AB system generated with \texttt{rebound} \protect\citep{rebound,reboundias15}. The plot coordinates are centered on the centre of mass of the system. The primary is as shown with a black star symbol and the secondary is shown as a black circle and its orbital path is visible as a red ellipse. The periastron of the orbit is denoted with a red dashed line. The inferior conjunction (transit) and superior conjunction (secondary eclipse) are both labelled with black arrows.}
    \label{fig:orbitplot}
\end{figure}

In addition, we find the secondary companion is on a highly eccentric orbit (e=\Be), which is plotted in Figure\,\ref{fig:orbitplot}, and a long period of \Bperiod\,days. The eccentricity of the companion's orbit, along with its mass, is consistent with the stellar regime described in \citet{Bowler2020mass-ecc} who found that high mass brown dwarfs and stellar companions show generally higher orbital eccentricities than low-mass brown dwarfs or planets -- showing a difference in formation mechanism. Since \OBJID\,B belongs to the stellar regime it is unlikely that it formed in a protoplanetary disk like a planet or low-mass brown dwarf. As Figures\,\ref{fig:period-mass}\,and\,\ref{fig:period-eccentricity} show, \OBJID\,B has one of the longest-period and most eccentric orbits of any low mass eclipsing binary companions known. The closest orbital separation of the binary is wide enough (at \peridistance\,AU) that we can expect that the evolution of \OBJID\,B is similar to that of a single star in a similar mass and radius range, with little effect from interaction with the primary star. This makes this system useful to test potential radius inflation mechanisms proposed for late M dwarf companions, this is discussed further in Section\,\ref{section:secondary_inflation}.

\begin{table}
    \centering
    \caption{\OBJID\,B parameters.}
    \begin{tabular}{c c c}
        \toprule
        Parameter&  Value & Source \\
        \hline
        \multicolumn{3}{c}{\textit{Fitted parameters}} \\ 
        \hline
        $\rtwo/\rone$& \Brr & \alles\ (\S\ref{section:allesfitter}) \\
        \noalign{\smallskip}
        $\left(\rone + \rtwo\right) / a$& \Brsuma & \alles\ (\S\ref{section:allesfitter}) \\
        \noalign{\smallskip}
        $\cos i$& \Bcosi & \alles\ (\S\ref{section:allesfitter}) \\
        \noalign{\smallskip}
        $T_{0;\text{tra}}$ (BJD)& \Bepoch & \alles\ (\S\ref{section:allesfitter}) \\
        $P_\text{orb}$ (days)& \Bperiod & \alles\ (\S\ref{section:allesfitter}) \\
        $\sqrt{e}\cos\omega$& \Bfc & \alles\ (\S\ref{section:allesfitter}) \\
        $\sqrt{e}\sin\omega$& \Bfs & \alles\ (\S\ref{section:allesfitter}) \\
        K (\kms)& \BK & \alles\ (\S\ref{section:allesfitter}) \\
        \hline
        \multicolumn{3}{c}{\textit{Derived parameters}} \\ 
        \hline
        \rtwo (\rjup)& \BRcompanionRjup & \alles\ (\S\ref{section:allesfitter}) \\
        \rtwo (\rsun)& \BRcompanionRsun & \alles\ (\S\ref{section:allesfitter}) \\
        \mtwo (\mjup)& \MBMjup & This work (\S\ref{section:mass_function}) \\
        \noalign{\smallskip}
        \mtwo (\msun)& \MBMsun & This work (\S\ref{section:mass_function}) \\
        \logg\ ($\log(\text{cgs})$) & $5.22\pm0.03$ & This work (\S\ref{section:results:occultation}) \\
        $a$ (AU)& \BaAU & \alles\ (\S\ref{section:allesfitter}) \\
        $r_\text{apastron}$ (AU) & \apodistance & This work* \\
        $r_\text{periastron}$ (AU) & \peridistance & This work* \\
        $b_\text{tra}$& \Bbtra & \alles\ (\S\ref{section:allesfitter}) \\
        \noalign{\smallskip}
        $b_\text{occ}$ & $0.12^{+0.09}_{-0.08}$ & This work (\S\ref{section:b_occ}) \\
        \noalign{\smallskip}
        $i (\deg)$ & \Bi & \alles\ (\S\ref{section:allesfitter}) \\
        $e$& \Be & \alles\ (\S\ref{section:allesfitter}) \\
        $\omega (\deg)$& \Bw & \alles\ (\S\ref{section:allesfitter}) \\
        \noalign{\smallskip}
        $T_{14}$ (hrs)& \BTtratot & (\S\ref{section:allesfitter}) \\
        \noalign{\smallskip}
        $T_{23}$ (hrs)& \BTtrafull & (\S\ref{section:allesfitter}) \\
        \noalign{\smallskip}
        $T_{58}$ (hrs)& \secondarydurtot & This work (\S\ref{section:occ_time}) \\
        $T_{67}$ (hrs)& \secondarydurfull & This work (\S\ref{section:occ_time}) \\
        $T_{0;\text{occ}}$ (BJD)& \Tnoughtocc & This work (\S\ref{section:occ_time}) \\
        $\phi_{0;\text{occ}}$ & \phinoughtocc & This work (\S\ref{section:occ_time}) \\
        \bottomrule
    \end{tabular}
    \begin{tablenotes}
      \small
      \item \textbf{*} The values $r_\text{apastron}=a(1+e)$ and $r_\text{periastron}=a(1-e)$ were calculated from the eccentricity, $e$, and semi-major axis, $a$, values returned from \alles.
    \end{tablenotes}
    \label{tab:secondary_params}
\end{table}

\subsection{Galactic kinematics and population membership}
\label{section:diskmember}
We computed $(U,V,W)$ galactic velocities using an updated version of the method from \citet{Johnson1987kinematics}. To do this, we used the parallax and proper motion values from \gaia\ DR3 (see Table\,\ref{tab:host_properties}) and calculated an error weighted mean of the instrumental radial velocity baselines shown in Table\,\ref{tab:ns_table}. This value is also shown in Table\,\ref{tab:host_properties} and is in reasonable agreement with the \gaia\,DR3 value. We also computed the Galactic total velocity, eccentricity and angular momentum of the orbit Z-component. The resulting values are all shown in Table~\ref{tab:host_properties}.

We used the velocities from Table~\ref{tab:host_properties} to determine the probability of membership for the thin disk, thick disk and halo using the framework from \citet{Bensby2003thinthickdisk} and the Galactic standard velocity dispersions from \cite{Bensby2003thinthickdisk, Bensby2014thinthickdisk, Reddy2006thinthickdisk, Chen2021thinthickdisk}. We also corrected  for the local standard of rest (LSR) using the values from \citet{Almeida-Fernandes2018kinematicage}. Lastly we combined the thin disk, thick disk and halo probabilities from the different velocity dispersions using a Bayesian Model Averaging method to calculate population membership probabilities.

The Bayesian analysis returned a probability of 80\% for thin disk membership, 20\% for thick disk and 0\% for the Galactic halo. While this indicates the star is likely a thin disk member, the above average total velocity $(v_\text{tot})$, eccentricity $(e_\text{gal})$ and angular momentum $(J_z)$ shown in Table~\ref{tab:host_properties} would usually imply a thick disk or halo star \citep{Yan2019}. We conclude that the most likely scenario is that \OBJID\ is a thin disk system that has undergone some kinematic heating in the recent past; perhaps through gravitational interactions with other stars.

Rossiter-McLaughlin \citep[RM;][]{Rossiter1924, McLaughlin1924, Triaud2018RM} measurements could be used to constrain the spin-orbit obliquity of the companion star, providing further clues to the dynamical history of the system and potential previous encounters with other stars. Using equation\,1 from \citet{Triaud2018RM}, we estimate an RM amplitude of $\sim$20\,m\,s$^{-1}$, which is detectable by current generation RV instruments such as \harps\ \citep{mayor2003harps}.

\subsection{Secondary eclipse}
\label{section:results:occultation}
Detection of a secondary eclipse / occultation of \OBJID\,B by the primary would allow for a more robust estimation of its luminosity and therefore temperature. Given the discrepancy between measured M dwarf temperatures and those predicted by models \citep{Spada2013radiusinflation, Parsons2018radiusinflation}, temperature measurements of detached late M dwarf companions such as \OBJID\,B via secondary eclipse or other means is vital to understanding this class of star as a whole.

\subsubsection{Inclination}
\label{section:b_occ}
To determine if the system is inclined such that a secondary eclipse could occur we calculated the impact parameter at occultation $b_\text{occ}$ using the formula laid out in \citet{Winn2014transits}. Using a Monte-Carlo approach to estimate errors we find $b_\text{occ}=0.12^{+0.09}_{-0.08}$, and hence we pedict that a secondary eclipse of \OBJID\,B should be detectable with sufficient photometric precision.

\subsubsection{Timing}
\label{section:occ_time}
We used the \texttt{contact\_points} function from the \texttt{pycheops} package \citep{Maxted2022pycheops} in order to calculate the predicted times of contact points during secondary eclipse $t_5, t_6, t_7$ and $t_8$ using the values fitted from our orbital solution (see Section\,\ref{section:allesfitter} and Table~\ref{tab:secondary_params}). We then used these to calculate the mid eclipse time ($T_{0;\text{occ}}$=\Tnoughtocc\,BJD), which corresponds to phase $\phi_{0;\text{occ}}$=\phinoughtocc. This would put a secondary eclipse in Sector 8 of the TESS data alongside the primary transit (although it cannot be seen in the data). We also calculate the full and total durations of occultation ($T_\text{tot;occ}=T_{58}$=\secondarydurtot\,hrs, $T_\text{full;occ}=T_{57}$=\secondarydurfull\,hrs,). The uncertainties were estimated via a monte-carlo approach and we quote the mean and standard deviation of our samples as the value and uncertainty for each parameter.

\subsubsection{Estimated depth and detectability}
\label{section:occ_depth}

We did not detect a secondary eclipse in any of the lightcurves obtained for \OBJID. A visual inspection of the lightcurves did not yield any detections and attempting to fit a secondary eclipse with \alles\ also returned a depth of 0. Since the inclination of the system is such that a secondary eclipse should occur this is most likely due to insufficient photometric precision. To test this we begun by estimating secondary eclipse depths of \OBJID\,B in various photometric bandpasses. 

We chose to test temperatures of the secondary between 2450K and 3050K as quoted for the M9V and M5V spectral types in \citet{2013PecautTeff}, which is the range we expect \OBJID\,B to lie in. We then used the \texttt{expecto}\footnote{Availiable at: \url{https://github.com/bmorris3/expecto}} package to fetch PHOENIX model spectra \citep{Husser2013specmodels} for the primary and secondary based on the respective values of \teff\ and \logg. The value of \logg\ for the secondary was calculated using the relation from \citet{2005SmalleyLogg} finding a value of $5.22\pm0.03\,\log(\text{cgs})$ We then use these model spectra along with various transmission filters and the fitted radius ratio to calculate the possible depth ranges shown in Table\,\ref{tab:occ_depths}.

\begin{table}
    \centering
    \caption{Estimated secondary eclipse depths for \OBJID\,B.}
    \begin{tabular}{c c}
	\toprule
	Bandpass & Depth (ppm)\\
	\hline
	\TESS\ & 108-370 \\
	\NGTS\ & 32-174 \\
	\CHEOPS\ & 60-224 \\
	\PLATO\ & 71-238 \\
	J & 539-1153 \\
	H &  685-1459\\
	K & 923-1879 \\
	\bottomrule
\end{tabular}
    \label{tab:occ_depths}
\end{table}

\begin{figure}
    \centering
    \includegraphics[width=\columnwidth]{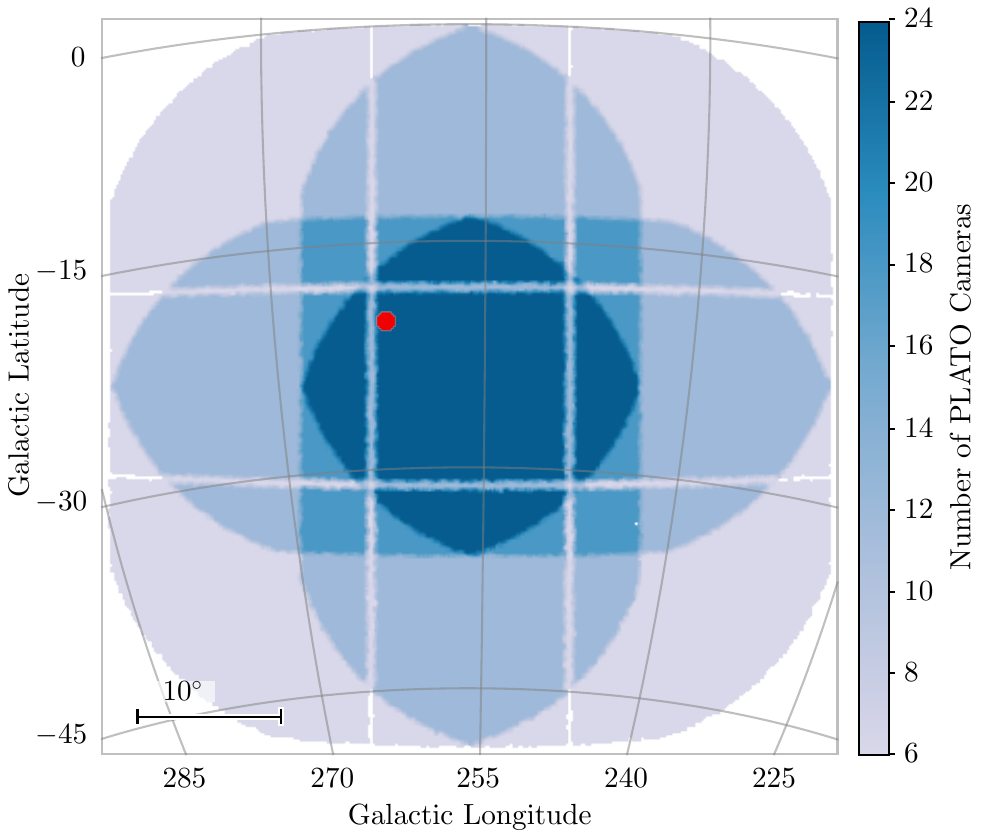}
    \caption{The PLATO LOPS2 field. Different shades of blue represent the areas where a star is observed with either 6, 12, 18 or 24 cameras, with darker shades indicating a greater number of cameras. \OBJID\ (marked with a red dot) is located in the dark blue region in the centre which will be observed with 24 cameras.}
    \label{fig:plato_field}
\end{figure}

The depths we find for \tess\ and \ngts\ are similar to the noise profiles for both instruments, which explains the non-detection of any secondary eclipse we see in all the photometric data for the target (see Section\,\ref{section:photobs}). We do, however, find significantly greater depths for the 2MASS J, H and K bandpasses indicating the eclipse would be detectable with these filters - although observing sufficient baseline to measure the depth accurately and precisely may be difficult from the ground given the expected long duration of eclipse. In addition we expect similar depths for \cheops\ and \plato, while both of these are similar depths as predicted for \tess\ and \ngts, the greater precision of \cheops\ and \plato\ may allow them to detect the secondary eclipse where \tess\ or \ngts\ may not be able to. \OBJID\ is in the P5 sample of the \plato\ \citep{Rauer2014Plato} Input Catalogue \citep[PIC;][]{Montalto2024PIC} and is situated near the centre of the \plato\ LOPS2 field and is hence observable with 24 cameras \citep[][]{Montalto2024PIC, Eschen2024Plato}. This is shown in Figure\,\ref{fig:plato_field}. The predicted noise is recorded in the \plato\ Input Catalogue as calculated using the \plato\ Instrument Noise Estimator \citep[PINE;][]{börner2024plato_noise}. Observing \OBJID\ with 24 \plato\ cameras is predicted to result in a precision of 101.8 ppm in one hour. This gives strong prospects of the secondary eclipse being detectable by \plato. With \plato's long baseline of at least two years (possibly 4) in LOPS2, 3-4 transits and/or secondary eclipses should be visible, allowing a building up of SNR with additional detections.

\subsection{Tidal circularisation}
\label{section:tcirc}

Tidal effects between binaries at large orbital separations such as \OBJID\ are expected to be extremely weak. We use the semi-empirical relations from \citet{1997ClaretTcirc} to estimate the tidal circularisation timescale of the system. We also estimate the timescale using the formulae laid out in \citet{Jackson2008tidalevolution}. These formulae do assume that the rotation of the primary and secondary stars is not synchronised, the eccentricity of the system is low and the orbital period is below 10\,days. Whilst the latter two assumptions are certainly not true for \OBJID\,AB, this can still provide a useful order of magnitude estimate and sanity check for the values estimated using the \citet{1997ClaretTcirc} method. From the \citet{1997ClaretTcirc} relation we find the circularisation timescale should be on the order of $10^6$ Gyr while the \citet{Jackson2008tidalevolution} relation returns values between $10^4-10^6$ Gyr depending on the chosen tidal quality factors. Both of these estimates are several orders of magnitude longer than the age of the universe, which leads us to conclude that the system will not circularise before the primary enters its red giant phase - at such time the system will likely become a common-envelope system. As well as the tidal circularization timescale, we also estimate the tidal synchronisation timescale using equation\,14 from \citet{Maxted2023EBLM}, finding a value of $\sim10^5-10^6$\,Gyr depending on rotational period. Again these lead us to conclude that tidal interactions in the system are incredibly weak and that no meaningful interaction between the primary and secondary will occur until the system reaches the common envelope stage.

\subsection{Radius inflation}
\label{section:secondary_inflation}

Many models attempting to explain the `radius-inflation' problem for M-dwarfs rely on magnetic effects \citep[e.g;][]{Chabrier2007mdwarfinflation, LopezMorales2007mdwarfinflation,Morales2022mdwarfinflation}. According to these models the inflation should be more prevalent in close binaries where interactions with the primary will `spin up' the secondary and lead to an enhanced magnetic dynamo. This faster rotation and coupled magnetic effects can inflate the radius of the secondary. However, it is currently unclear whether radius inflation is limited to only short-period EBLMs. \citet{vonBoetticher2019EBLM5} found no link between orbital period and radius inflation while \citet{Swayne2024EBLM11} found inflation was greater for M dwarfs in short-period binaries. However, these studies were each based on a relatively small number of EBLM systems, all with considerably shorter orbital periods than \OBJID\,B. Given its position in the extreme long-period region of parameter space shown in Figures\,\ref{fig:period-mass}\,and\,\ref{fig:period-eccentricity}, the potential inflation of \OBJID\,B has strong ramifications to these magnetic models attempting to explain inflation.

\OBJID\,B appears to have a radius $\sim1.0\%$ above what the nearest point on the isochrone models from \citet{Baraffe2015Mdwarfmodels} would predict (see Figure\,\ref{fig:mass-radius}). This is consistent with \citet{Maxted2023EBLM} who conclude that radius inflation for the lowest-mass M-dwarfs is $\sim2\%$. However, we find this difference to be statistically insignificant as the vector difference between the mass and radius of \OBJID\,B and the nearest point of the isochrone is $0.97\sigma$. While more precise mass and radius measurements could increase this, given the already relatively precise mass and radius of \OBJID\,B, we find it highly unlikely that this discrepancy is statistically significant and thus is unlikely to be astrophysical in nature. It should also be noted that the inflation is model dependent, and comparing the mass and radius of \OBJID\,B to another evolutionary model with different input physics (e.g. the equation of state) will change the apparent inflation. Filling out the long-period parameter space could allow for a population level determination of radius inflation as a function of orbital period, but this is currently difficult while the number of extremely long-period systems remains low.

\section{Summary and Conclusion}
\label{section:conclusion}
The \OBJID\,AB system is a long-period and highly eccentric EBLM system with an evolved G-type primary and late M-dwarf secondary near the hydrogen burning limit. We expect tidal effects on both the primary and secondary to be negligible due to the relatively wide orbital separation of both stars at periastron, despite the high eccentricity.

We first discovered evidence of a secondary companion via a single transit detection in \tess\ Sector 8 with a depth consistent with a Jupiter-sized planet and an additional transit in Sector 29 which we extracted from the data despite initially being hidden due to scattered Earth light. We then began a campaign of spectroscopic and photometric follow-up observations. An observation of a transit egress with \ngts\ was able to confirm the orbital period as \Bperiod\,days. Radial velocity measurements from \coralie\ and \harps\ spectra revealed the companion was stellar in mass and on a highly eccentric orbit. The spectra used to measure these radial velocities were also used to measure atmospheric parameters of the primary star \OBJID\,A, revealing it to be an evolved G sub-giant.

Global analysis of the photometric and radial velocity data provided a complete orbital solution for the system with an orbital period as \Bperiod\,days and an eccentricity of \Be, making \OBJID\,AB one of the longest period and most eccentric EBLMs known. This allowed us to measure a mass of \MBMjup\,\mjup\,and a radius of \BRcompanionRjup\,\rjup, which is near the hydrogen limit ($\sim$80\,\mjup). We compared this against evolutionary models from the literature and found a difference of $1.26\sigma$ for the radius, which indicates \OBJID\,B is unlikely to be inflated. More precise mass and radius measurements could refine this difference and potentially find a significant deviation but this is unlikely.

Longer-period M dwarf secondary companions are further away from their primaries allowing their evolution to be reasonably approximated as analogous to single star evolution for a similar spectral type, which makes them useful for calibrating evolutionary models for single M dwarfs. The \OBJID\,AB system provides a valuable benchmark for comparison against such models, with well constrained parameters including stellar age due to the evolved nature of the primary. Further RV measurements, focused around periastron, could further constrain the mass of \OBJID\,B, allowing for an accurate determination of whether its radius is inflated. In addition, \OBJID\ is located near the centre of the proposed southern LOPS2 field for the upcoming \plato\ mission \citep{Rauer2014Plato}. With the longer observation baseline and greater photometric precision of \plato\ it would be possible to detect additional transits of \OBJID\,B and further refine its radius as well as detecting its secondary eclipse and potentially even rotational modulation of \OBJID\,A.

As the \tess\ mission progresses and additional observatories with longer observation baselines such as \plato\ become available, more discoveries of bright nearby long-period EBLMs will become possible. This will allow the region of parameter space currently occupied by \OBJID\,B and a few other systems to become more populated - allowing for population level analysis of M-dwarf inflation at these long-period regimes.

\section*{Acknowledgements}

We would like to thank the assistant editor and anonymous reviewer for their comments that helped to improve this paper.

This paper includes data collected by the \TESS\ mission. Funding for the \TESS\ mission is provided by the NASA Explorer Program. Resources supporting this work were provided by the NASA High-End Computing (HEC) Program through the NASA Advanced Supercomputing (NAS) Division at Ames Research Center for the production of the SPOC data products. The TESS team shall assure that the masses of fifty (50) planets with radii less than 4\rearth\ are determined.


We acknowledge the use of public \TESS\ Alert data from pipelines at the \TESS\ Science Office and at the \TESS\ Science Processing Operations Center.


This paper includes data collected by the TESS mission that are publicly available from the Mikulski Archive for Space Telescopes (MAST).

This work is based in part on data collected under the NGTS project at the ESO La Silla Paranal Observatory. The NGTS facility is operated by a consortium of institutes with support from the UK Science and Technology Facilities Council (STFC) under projects ST/M001962/1, ST/S002642/1 and ST/W003163/1.

TR is supported by an STFC studentship. C.A.W. and E.dM. would like to acknowledge support from the UK Science and Technology Facilities Council (STFC, grant number ST/X00094X/1). JSJ gratefully acknowledges support by FONDECYT grant 1240738 and from the ANID BASAL project FB210003.
This work has been carried out within the framework of the NCCR PlanetS supported by the Swiss National Science Foundation under grants 51NF40\_182901 and 51NF40\_205606.
EG gratefully acknowledges support from the UK Science and Technology Facilities Council (STFC; project reference ST/W001047/1). 

R.B. acknowledges support from FONDECYT Project 1241963 and from ANID -- Millennium  Science  Initiative -- ICN12\_009.

T.T. acknowledges support by the BNSF program "VIHREN-2021" project No. KP-06-DV/5.

For the purpose of open access, the author has applied a Creative Commons Attribution (CC BY) licence to the Author Accepted Manuscript version arising from this submission.

\section*{Data Availability}

The \tess\ data is accessible via the MAST (Mikulski Archive for Space Telescopes) portal at \url{https://mast.stsci.edu/portal/Mashup/Clients/Mast/Portal.html}.




The \harps\ data can be obtained from the ESO archive at \url{http://archive.eso.org/cms.html}

Any code used for analysis or in producing the plots in this paper can be made available upon reasonable request to the author(s).


\bibliographystyle{mnras}
\bibliography{refs}



\appendix

\section{\alles\ results tables}
\label{section:allesfitter_tables}

\begin{table*}
    \centering
    \caption{\alles\ global model priors and fitted parameters.}
    \begin{tabular}{c c c c c}
	\toprule
	Parameter & Initial guess & Prior & Fitted value & units\\
	\hline
	$\rtwo / \rone$&0.0893259057224787&$\mathcal{U}\left(0.07,0.14\right)$&\Brr&\\
    \noalign{\smallskip}
	$(\rone + \rtwo) / a_\mathrm{B}$&0.0113578760330941&$\mathcal{U}\left(0.008,0.05\right)$&\Brsuma&\\
    \noalign{\smallskip}
	$\cos{i_\mathrm{B}}$&0.0057874459294714&$\mathcal{U}\left(0,1.0\right)$&\Bcosi&\\
    \noalign{\smallskip}
	$T_{0;\mathrm{B}}$&2459486.073495738&$\mathcal{U}\left(2459485.5,2459486.5\right)$&\Bepoch&bjd\\
	$P_\mathrm{B}$&193.3585642579528&$\mathcal{U}\left(192.0,194.0\right)$&\Bperiod&d\\
	$\sqrt{e_\mathrm{B}} \cos{\omega_\mathrm{B}}$&-0.835487&$\mathcal{U}\left(-1.0,1.0\right)$&\Bfc&\\
	$\sqrt{e_\mathrm{B}} \sin{\omega_\mathrm{B}}$&0.131535&$\mathcal{U}\left(-1.0,1.0\right)$&\Bfs&\\
	$K_\mathrm{B}$&4.651554575593812&$\mathcal{U}\left(0.0,12.0\right)$&\BK&$\mathrm{km/s}$\\
	dilution TESS8&0.02864754&Fixed&&\\
	dilution TESS29&0.0140216899999999&Fixed&&\\
	$q_{1; \mathrm{NGTS}}$&0.37&$\mathcal{N}\left(0.37,0.01\right)$&\hostldcqoneNGTS&\\
	$q_{1; \mathrm{NGTS}}$&0.40&$\mathcal{N}\left(0.40,0.08\right)$&\hostldcqtwoNGTS&\\
	$q_{1; \mathrm{TESS8}}$&0.315&$\mathcal{N}\left(0.315,0.008\right)$&\hostldcqoneTESSeight&\\
	$q_{2; \mathrm{TESS8}}$&0.39&$\mathcal{N}\left(0.39,0.08\right)$&\hostldcqtwoTESSeight&\\
	$q_{1; \mathrm{TESS29}}$&0.315&$\mathcal{N}\left(0.315,0.008\right)$&\hostldcqoneTESStwentynine&\\
	$q_{2; \mathrm{TESS29}}$&0.39&$\mathcal{N}\left(0.39,0.08\right)$&\hostldcqtwoTESStwentynine&\\
	$\log{\sigma_\mathrm{TESS8}}$&-7.31&$\mathcal{N}\left(-7.31,0.04\right)$&\lnerrfluxNGTS&$\log({ \mathrm{rel. flux.}) }$\\
	$\log{\sigma_\mathrm{TESS29}}$&-6.54&$\mathcal{N}\left(-6.54,0.02\right)$&\lnerrfluxTESSeight&$\log({ \mathrm{rel. flux.}) }$\\
	$\log{\sigma_\mathrm{NGTS}}$&-7.4&$\mathcal{N}\left(-7.4,0.7\right)$&\lnerrfluxTESStwentynine&$\log({ \mathrm{rel. flux.}) }$\\
	$\log{\sigma_\mathrm{CORALIE}}$&-5.0&$\mathcal{U}\left(-7.6,-1.6\right)$&\lnjitterrvCORALIE&$\log(\mathrm{km/s})$\\
    \noalign{\smallskip}
	$\log{\sigma_\mathrm{HARPS}}$&-6.0&$\mathcal{U}\left(-7.6,-1.6\right)$&\lnjitterrvHARPS&$\log(\mathrm{km/s})$\\
    $\log{\sigma_\mathrm{FEROS}}$&-5.0&$\mathcal{U}\left(-7.6,-1.6\right)$&\lnjitterrvFEROS&$\log(\mathrm{km/s})$\\
	GP $\ln \sigma$ (TESS8)&-7.7&$\mathcal{N}\left(-7.7,0.4\right)$&\baselinegpmaternthreetwolnsigmafluxTESSeight&\\
	GP $\ln \rho$ (TESS8)&0.3&$\mathcal{N}\left(0.3,0.6\right)$&\baselinegpmaternthreetwolnrhofluxTESSeight&\\
	GP $\ln \sigma$ (TESS29)&-7.01&$\mathcal{N}\left(-7.01,0.08\right)$&\baselinegpmaternthreetwolnsigmafluxTESStwentynine&\\
	GP $\ln \rho$ (TESS29)&-1.79&$\mathcal{N}\left(-1.7,0.2\right)$&\baselinegpmaternthreetwolnrhofluxTESStwentynine&\\
	offset CORALIE&78.71&$\mathcal{N}\left(78.71,0.95\right)$&\baselineoffsetrvCORALIE&$\mathrm{km/s}$\\
	offset HARPS&78.71&$\mathcal{N}\left(78.71,0.95\right)$&\baselineoffsetrvHARPS&$\mathrm{km/s}$\\
    offset FEROS&78.71&$\mathcal{N}\left(78.71,0.95\right)$&\baselineoffsetrvFEROS&$\mathrm{km/s}$\\
	\bottomrule
\end{tabular}
    \label{tab:ns_table}
\end{table*}

\begin{table}
    \centering
    \caption{System derived parameters.}
    \begin{tabular}{c c}
    \toprule
    Derived parameter & Value \\ 
    \hline 
    Host radius over semi-major axis B; $\rone/a_\mathrm{B}$ & \BRstarovera \\ 
    \noalign{\smallskip}
    Semi-major axis B over host radius; $a_\mathrm{B}/\rone$ & \BaoverRstar \\
    \noalign{\smallskip}
    Companion radius B over semi-major axis B; $\rtwo/a_\mathrm{B}$ & \BRcompanionovera \\
    Companion radius B; $\rtwo$ ($\mathrm{R_{\oplus}}$) & \BRcompanionRearth \\
    Companion radius B; $\rtwo$ ($\mathrm{R_{jup}}$) & \BRcompanionRjup \\ 
    Semi-major axis B; $a_\mathrm{B}$ ($\mathrm{R_{\odot}}$) & \BaRsun \\ 
    Semi-major axis B; $a_\mathrm{B}$ (AU) & \BaAU \\ 
    Inclination B; $i_\mathrm{B}$ (deg) & \Bi \\ 
    Eccentricity B; $e_\mathrm{B}$ & \Be \\ 
    Argument of periastron B; $w_\mathrm{B}$ (deg) & \Bw \\
    \noalign{\smallskip}
    Mass ratio B; $q_\mathrm{B}$ & \Bq\\
    \noalign{\smallskip}
    Companion mass B; $\mtwo$ ($\mathrm{M_{\oplus}}$) & \BMcompanionMearth \\
    \noalign{\smallskip}
    Companion mass B; $\mtwo$ ($\mathrm{M_{jup}}$) & \BMcompanionMjup \\
    \noalign{\smallskip}
    Companion mass B; $\mtwo$ ($\mathrm{M_{\odot}}$) & \BMcompanionMsun \\
    \noalign{\smallskip}
    Impact parameter B; $b_\mathrm{tra;B}$ & \Bbtra \\
    \noalign{\smallskip}
    Total transit duration B; $T_\mathrm{tot;B}$ (h) & \BTtratot \\
    \noalign{\smallskip}
    Full-transit duration B; $T_\mathrm{full;B}$ (h) & \BTtrafull \\ 
    \noalign{\smallskip}
    Companion density B; $\rho_\mathrm{B}$ (cgs) & \Bdensity \\ 
    \noalign{\smallskip}
    Companion surface gravity B; $g_\mathrm{B}$ (cgs) & \Bsurfacegravity \\ 
    Equilibrium temperature B; $T_\mathrm{eq;B}$ (K) & \BTeq \\ 
    Combined host density from all orbits; $\rho_\mathrm{A; combined}$ (cgs) & \Bhostdensity \\
    \bottomrule
\end{tabular}
    \label{tab:ns_derived_table1}
\end{table}

\begin{table}
    \centering
    \caption{Instrumental derived parameters.}
    \begin{tabular}{c c}
    \toprule
    Derived parameter & Value \\ 
    \hline 
    Transit depth (undil.) B; $\delta_\mathrm{tr; undil; B; TESS8}$ (ppt) & \BdepthtrundilTESSeight \\
    \noalign{\smallskip}
    Transit depth (dil.) B; $\delta_\mathrm{tr; dil; B; TESS8}$ (ppt) & \BdepthtrdilTESSeight \\ 
    Transit depth (undil.) B; $\delta_\mathrm{tr; undil; B; TESS29}$ (ppt) & \BdepthtrundilTESStwentynine \\ 
    Transit depth (dil.) B; $\delta_\mathrm{tr; dil; B; TESS29}$ (ppt) & \BdepthtrdilTESStwentynine \\ 
    Transit depth (undil.) B; $\delta_\mathrm{tr; undil; B; NGTS}$ (ppt) & \BdepthtrundilNGTS \\
    \noalign{\smallskip}
    Transit depth (dil.) B; $\delta_\mathrm{tr; dil; B; NGTS}$ (ppt) & \BdepthtrdilNGTS \\ 
    Limb darkening; $u_\mathrm{1; TESS8}$ & \hostldcuoneTESSeight \\ 
    Limb darkening; $u_\mathrm{2; TESS8}$ & \hostldcutwoTESSeight \\ 
    Limb darkening; $u_\mathrm{1; TESS29}$ & \hostldcuoneTESStwentynine \\ 
    Limb darkening; $u_\mathrm{2; TESS29}$ & \hostldcutwoTESStwentynine \\ 
    Limb darkening; $u_\mathrm{1; NGTS}$ & \hostldcuoneNGTS \\ 
    Limb darkening; $u_\mathrm{2; NGTS}$ & \hostldcutwoNGTS \\ 
    \bottomrule
\end{tabular}
    \label{tab:ns_derived_table2}
\end{table}

\section{\harps\ masked regions}
\label{section:appendix_bumps}

\begin{table}
    \centering
    \caption{\harps\ masked wavelength ranges.}
    \begin{tabular}{cc}
	\toprule
	Wavelength range (nm) & Wavelength range (nm)\\
	\hline
	390.40-390.50 & 488.87-489.12\\
	390.99-391.16 & 492.84-493.15\\
	392.26-392.33 & 498.14-498.43\\
	392.83-392.85 & 508.68-509.00\\
	395.94-396.12 & 514.43-514.75\\
	399.27-399.49 & 519.40-519.45\\
	402.60-402.78 & 522.87-522.91\\
	405.87-406.06 & 529.83-529.87\\
	408.94-408.95 & 534.45-534.50\\
	409.10-409.30 & 545.10-545.16\\
	411.10-411.15 & 552.15-552.37\\
	418.50-418.75 & 557.70-557.75\\
	422.15-422.35 & 562.46-562.50\\
	425.74-425.92 & 566.89-567.14\\
	431.98-432.02 & 582.38-582.43\\
	433.78-433.80 & 582.60-583.05\\
	435.05-435.08 & 583.34-583.38\\
	436.10-436.40 & 598.04-598.50\\
	438.16-438.18 & 599.30-599.34\\
	440.25-440.05 & 599.98-600.70\\
	441.31-441.35 & 601.83-602.34\\
	444.29-444.35 & 603.85-604.34\\
	444.51-444.55 & 605.99-606.62\\
	446.59-446.63 & 607.88-608.30\\
	447.75-447.80 & 609.17-610.38\\
	451.05-451.10 & 618.50-619.30\\
	455.97-456.26 & 629.84-630.20\\
	456.31-456.35 & 637.16-637.20\\
	460.32-460.70 & 637.58-637.60\\
	460.80-460.85 & 647.91-648.05\\
	473.37-473.70 & 652.19-652.22\\
	478.31-478.46 & 676.99-677.03\\
	483.06-483.25 & \\
	\bottomrule
 \end{tabular}
    \label{tab:bumps}
\end{table}

\section{\alles\ plots}
\label{section:allesfitter-plots}

\begin{figure*}
    \centering
    \includegraphics[width=\textwidth]{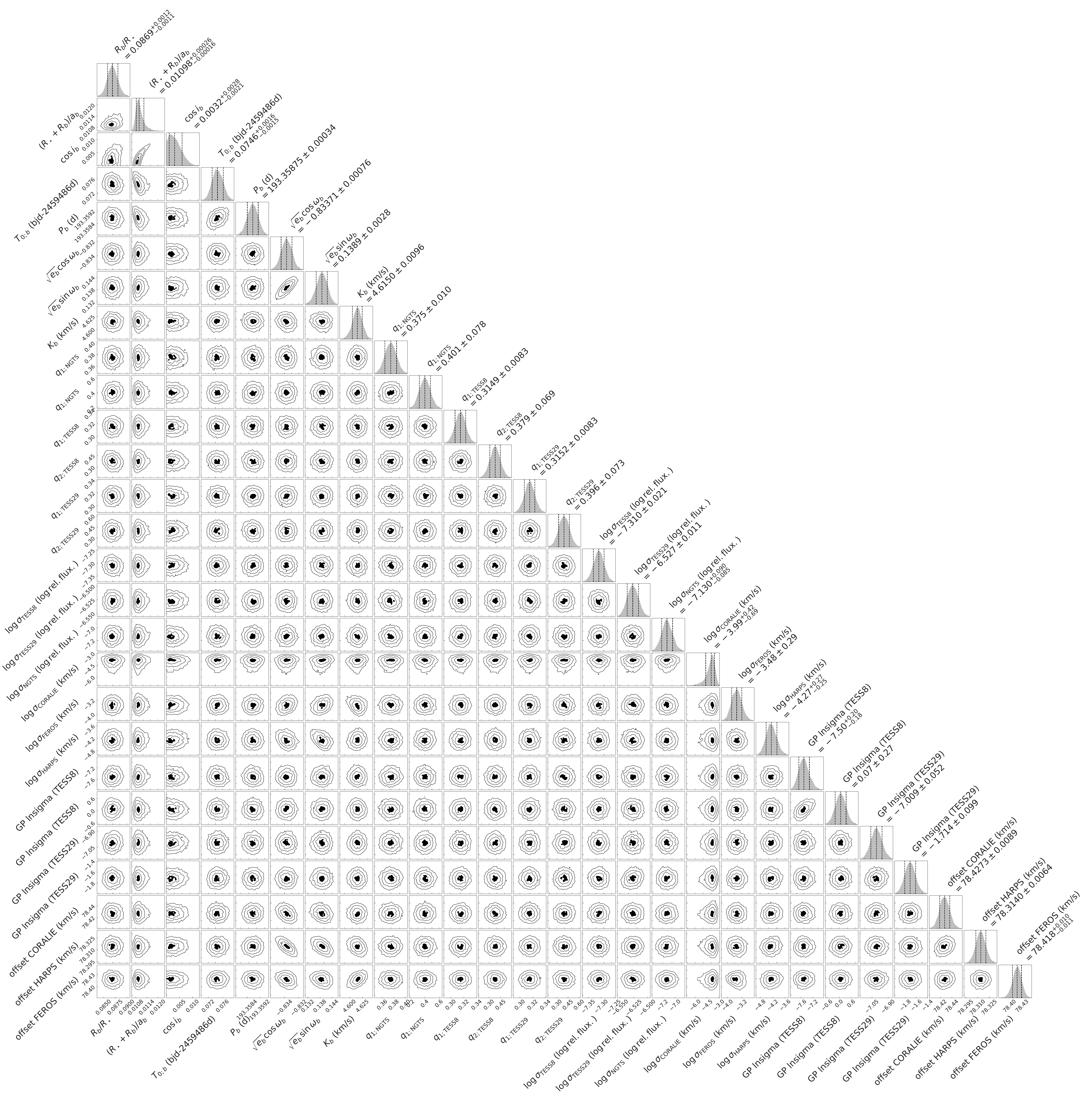}
    \caption{Corner plot of fitted parameters from the \alles\ global model.}
    \label{fig:allesfitter_corner}
\end{figure*}

\begin{figure*}
    \centering
    \includegraphics[width=\textwidth]{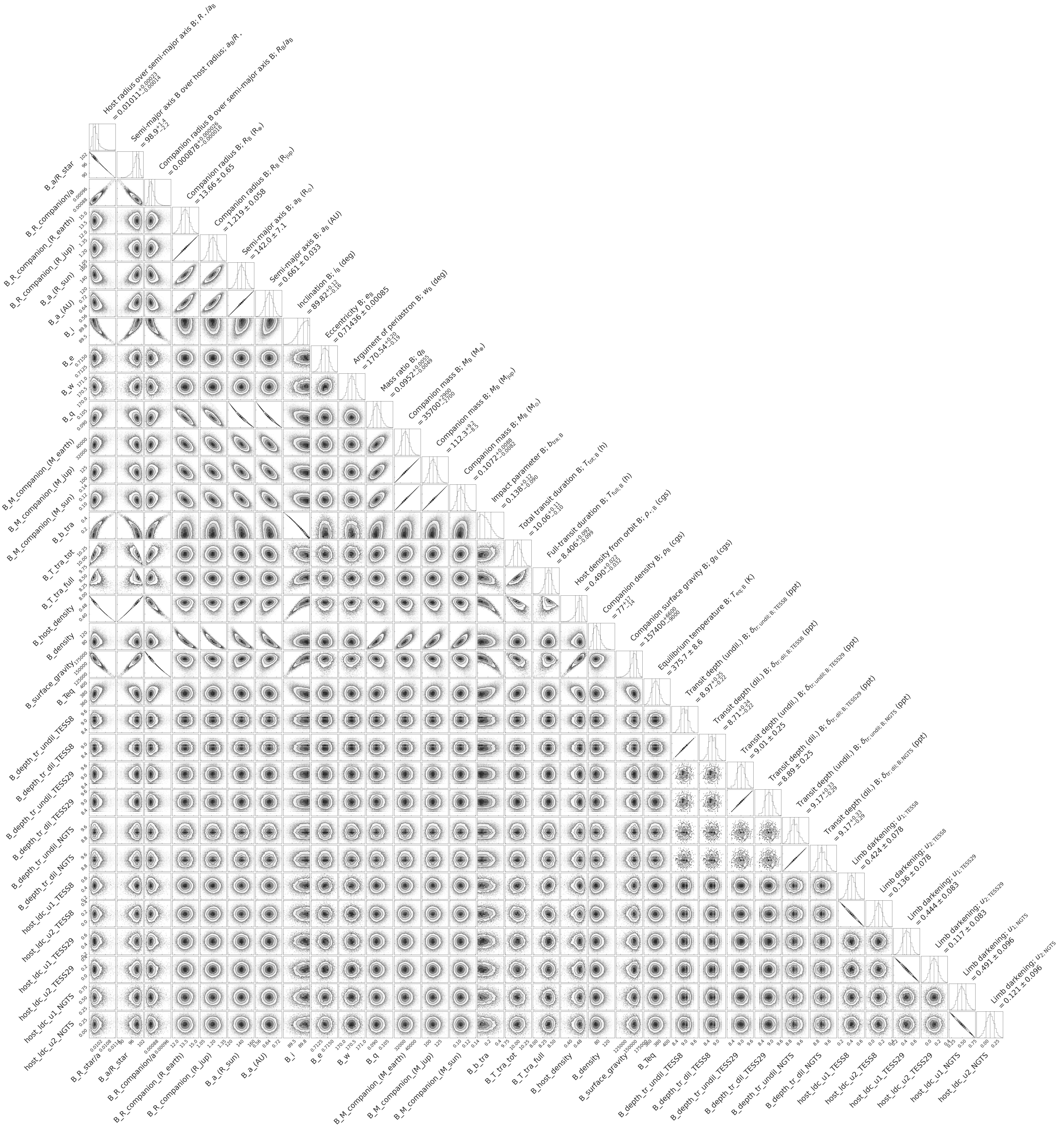}
    \caption{Corner plot of derived parameters from the \alles\ global model.}
    \label{fig:allesfitter_derived_corner}
\end{figure*}

\section{\isochrones\ plots}
\label{section:isochrones-plots}

\begin{figure*}
    \centering
    \includegraphics[width=\textwidth]{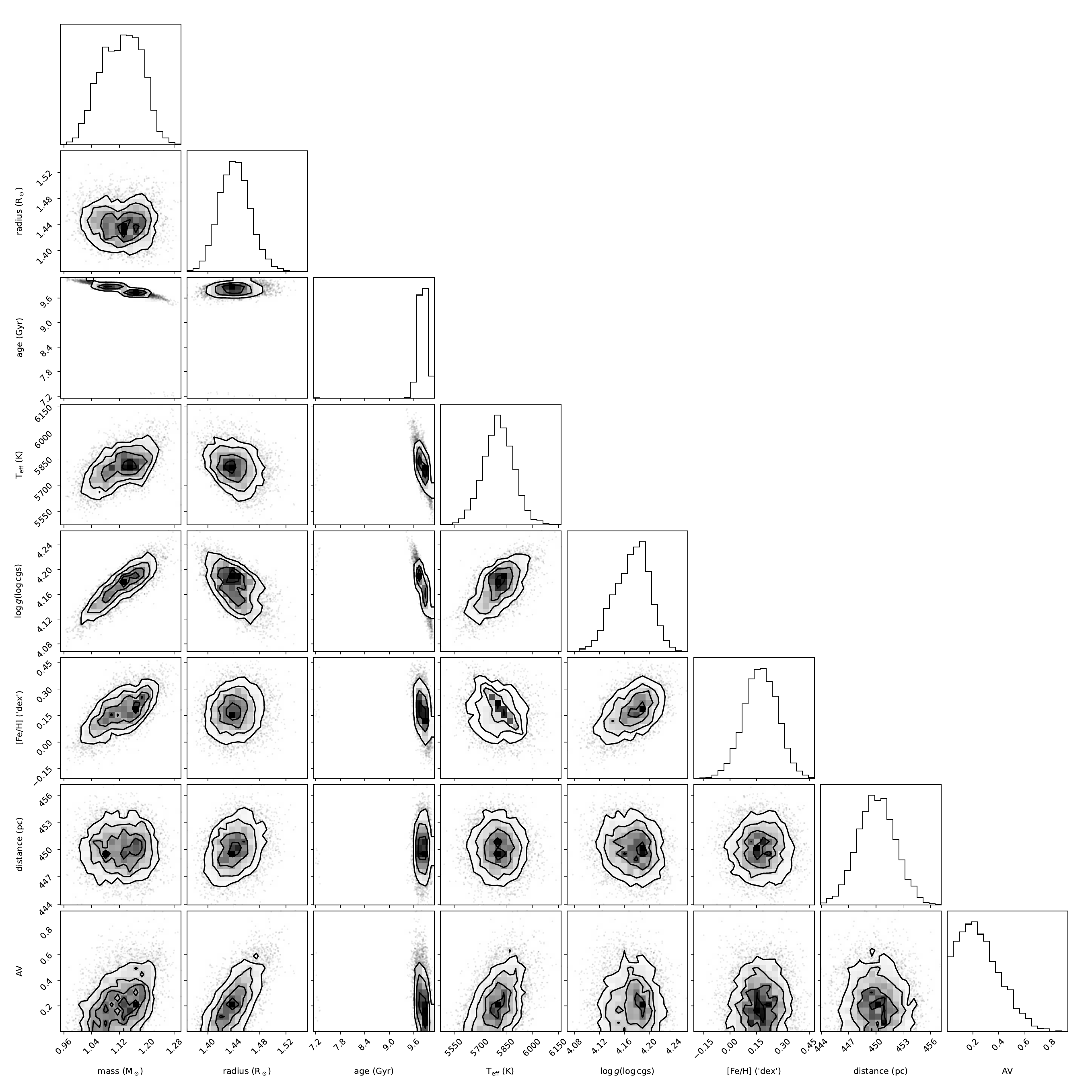}
    \caption{Corner plot of physical parameters from \isochrones\ using priors from \paws.}
    \label{fig:isochrones_corner_physical}
\end{figure*}

\begin{figure*}
    \centering
    \includegraphics[width=\textwidth]{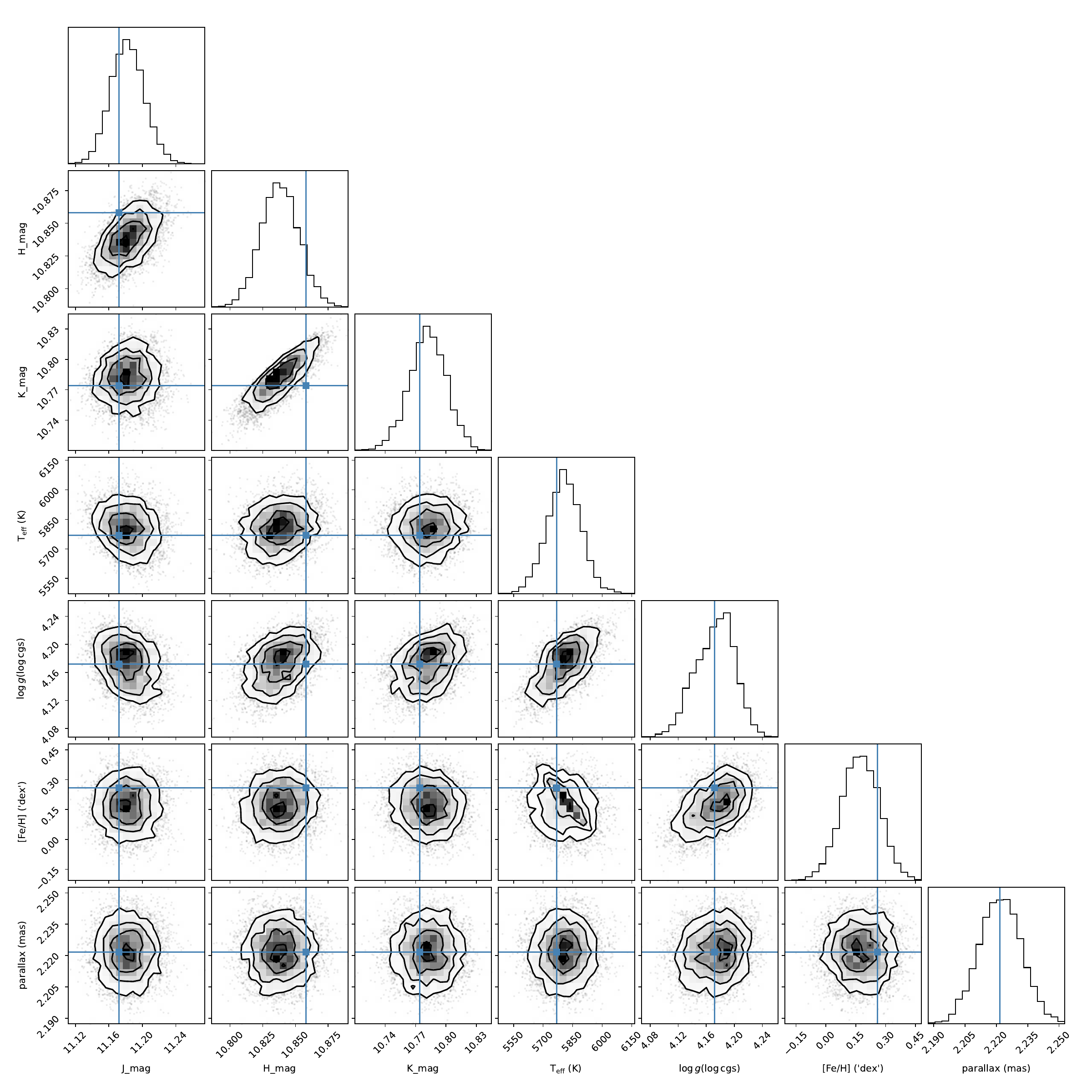}
    \caption{Corner plot of observables from \isochrones\ using priors from \paws.}
    \label{fig:isochrones_corner_observed}
\end{figure*}

\section{\tess\ raw lightcurves}
\label{section:bigtessplot}

\begin{figure*}
    \centering
    \includegraphics[width=\textwidth]{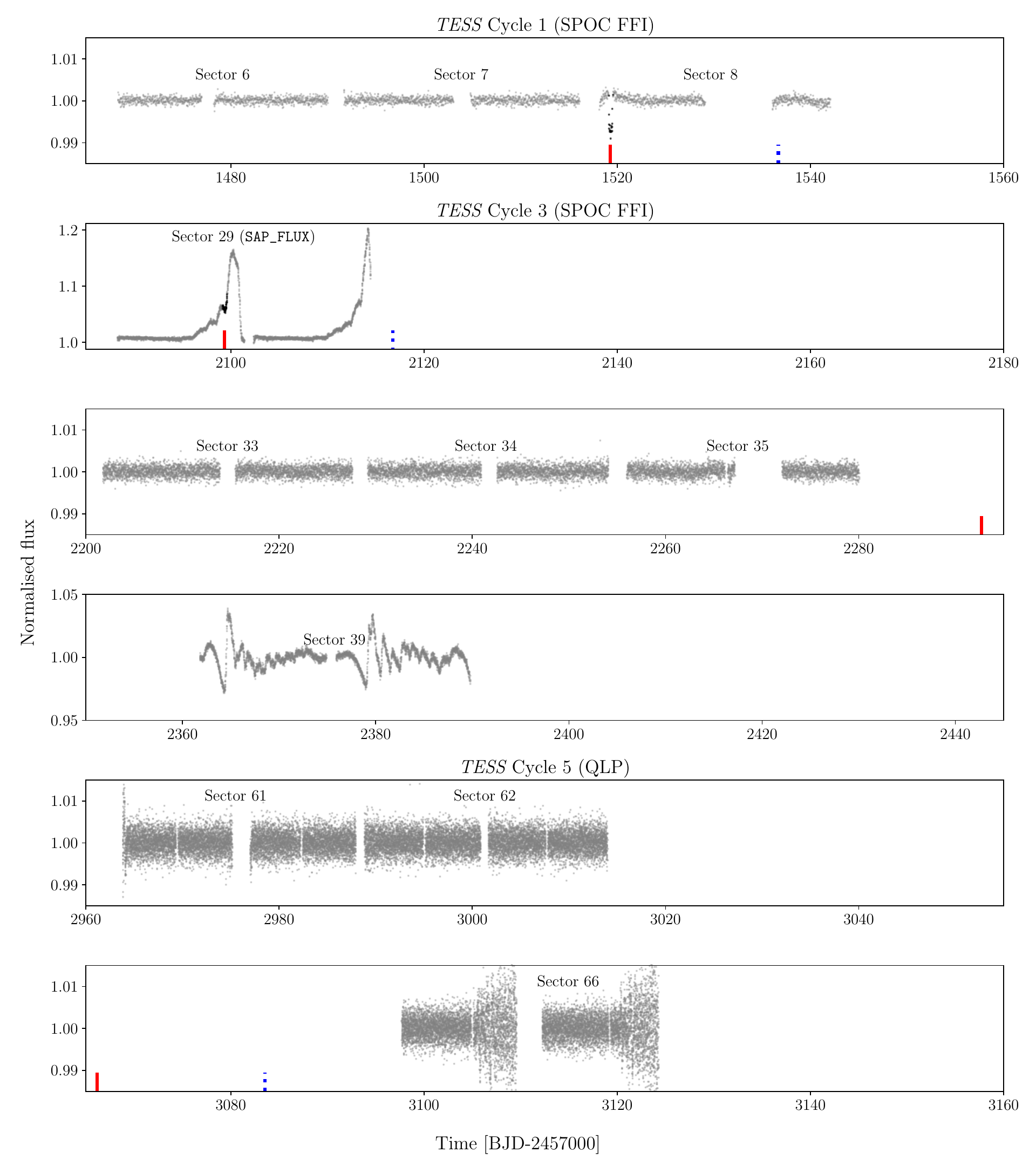}
    \caption{All \tess\ lightcurves for \OBJID\ plotted in grey with transits marked by red vertical lines, secondary eclipses with blue dotted vertical lines and in-transit data-points are highlighted black. Each panel shows a time-span of 95 days so that all the transits happening within each \tess\ cycle can be seen. The top panel represents Cycle 1, the middle three show Cycle 3 and the bottom two Cycle 5. For Sectors 6-8 and 33-39 we plot the \texttt{PDCSAP\_FLUX} flux from the SPOC FFI lightcurves while for Sector 29 we plot the \texttt{SAP\_FLUX}. The lightcurves for sectors 61-66 are sourced from the Quick Look Pipeline \citep[QLP;][]{huang2020qlp} and the detrended \texttt{DET\_FLUX}.}
    \label{fig:all_tess}
\end{figure*}

\section{The \ngts-EB naming convention}
\label{section:acronym}

We have created the \ngts-EB naming system to describe eclipsing binary systems discovered by the Next Generation Transit Survey \citep[\ngts;][]{wheatley2018ngts}. This naming system has been registered with the International Astronomical Union (IAU) Dictionary of Nomenclature of Celestial Objects \footnote{Available at: \url{https://cds.unistra.fr//cgi-bin/Dic?NGTS-EB}}. The Acronym \ngts-EB stands for Next Generation Transit Survey - Eclipsing Binary and the naming convention follows the format \ngts-EB-NNN\,A where NNN is a sequential number given to each system and A and B denote the individual components of the system with A typically being the more massive component. We designate the system that is the focus of this paper \ngts-EB-7. Despite being the first object to receive an \ngts-EB moniker, we give it the number 7 as six previous eclipsing binary systems have been discovered by \ngts, which we retroactively give \ngts-EB numbers as shown in Table\,\ref{tab:ngts-eb}.

\begin{table*}
    \centering
    \caption{\ngts-EB systems.}
    \begin{tabular}{cccccccc}
        \toprule
        \ngts-EB \#&  Original designation &TICID &\gaia\,DR3 ID &2MASS ID &RA &DEC &Reference\\
        \hline
        1& \ngts\,J052218.2-250710.4& 31054255& 2957804068198875776& J05221817-2507112& $5^\text{h}22^\text{m}18.17^\text{s}$& $-25\degree07{'}10.80^{''}$&1\\
        2& \ngts\,J214358.5-380102& 197570458& 6586032117320121856& J21435859-3801027& $21^\text{h}43^\text{m}58.59^\text{s}$& $-38\degree1{'}3.41^{''}$&2\\
        3& TIC\,231005575& 231005575& 4912474299133826560& J01400127-5431218& $1^\text{h}40^\text{m}1.35^\text{s}$& $-54\degree31{'}21.99^{''}$&3\\
        4& \ngts\,J0930-18& 176772671 & 5678383069566263552& J09301604-1850353& $9^\text{h}30^\text{m}16.00^\text{s}$& $-18\degree50{'}35.03^{''}$&4\\
        5& \ngts\,J0002-29& 313934158& 2320868389659322368& J00024841-2953539& $0^\text{h}2^\text{m}48.45^\text{s}$& $-29\degree53{'}53.89^{''}$& 5\\
        6& TIC\,320687387& 320687387& 6641131183310690432& J19511834-5532469& $19^\text{h}51^\text{m}18.41^\text{s}$& $-55\degree32{'}47.50^{''}$&6\\
        7& \OBJID & 238060327& 5504617415848984320& J06564704-5204263& $6^\text{h}56^\text{m}47.04^\text{s}$& $-52\degree04{'}26.11^{''}$& This work\\
        \bottomrule
    \end{tabular}
    \begin{tablenotes}
      \small
      \item \textbf{References:} 1:\cite{2018CasewellNGTSEB}, 2:\cite{2020ActonNGTSEB1}, 3:\cite{Gill2020monofind}, 4: \cite{2020ActonNGTSEB2}, 5:\cite{2021SmithNGTSEB}, 6:\cite{2022GillNGTSEB}. 
    \end{tablenotes}
    \label{tab:ngts-eb}
\end{table*}


\bsp	
\label{lastpage}
\end{document}